\documentclass[preprintnumbers,amsmath,amssymb,floatfix,10pt,prd,onecolumn,superscriptaddress,nofootinbib]{revtex4-2}
\usepackage{latexsym}
\usepackage{epsfig}
\usepackage{epstopdf}
\usepackage{graphicx}
\usepackage{amssymb}
\usepackage{amsmath}
\usepackage{mathtools}
\usepackage{amsfonts}
\usepackage{subfigure}
\usepackage{dcolumn}
\usepackage{bm}
\usepackage{color}
\usepackage{comment}
\usepackage[utf8]{inputenc}
\usepackage[dvipsnames]{xcolor}
\usepackage{hyperref}
\hypersetup{colorlinks=true,linkcolor=blue,citecolor=red,urlcolor=magenta}
\usepackage{orcidlink}

\begin{document}
\title{\bf On optical properties of rotating wormhole in Bopp-Podolsky electrodynamics}
\author{Muhammad Ali Raza \orcidlink{0009-0001-1281-8702}}
\email{maliraza01234@gmail.com}
\affiliation{Department of Mathematics, COMSATS University Islamabad, Lahore Campus, Lahore, Pakistan}
\author{Francisco Tello-Ortiz \orcidlink{0000-0002-7104-5746}}
\email{francisco.tello@ufrontera.cl}
\affiliation{Departamento de Ciencias Físicas, Universidad de La Frontera, Casilla 54-D, 4811186 Temuco, Chile.}
\author{M. Zubair \orcidlink{0000-0003-2227-788X}}
\email{mzubairkk@gmail.com;drmzubair@cuilahore.edu.pk}
\affiliation{National Astronomical Observatories, Chinese Academy of Sciences, Beijing 100101, China}
\affiliation{Department of Mathematics, COMSATS University Islamabad, Lahore Campus, Lahore, Pakistan}
\author{Y. Gómez-Leyton}
\email{y.gomez@ucn.cl}
\affiliation{Departamento de Física, Universidad Católica del Norte, Av. Angamos 0610, Antofagasta, Chile.}

\begin{abstract}
In this work, we consider a static wormhole in Bopp-Podolsky electrodynamics and convert it into its rotating counterpart by reducing it into Morris-Thorne form. We further study the null geodesics and effective potential along with the shadows for inner and outer unstable orbits for specific choices of parameters. It is found that for some cases smooth shadow curves are formed and for a few cases, the shadows formed are cuspy. All parameters have a significant impact on the shadows except for the parameter $b$ when either $a$ or $Q$ are kept small. We also analyze the gravitational lensing in the strong regime, considering that the observer and the source are on opposite sides of the throat. For this situation, we explore in detail the behavior of the deflection angle, Einstein rings and lensing observables.
\end{abstract}
\maketitle
\date{\today}

\section{Introduction}\label{int}
Despite its success, Maxwell's linear electrodynamics \cite{1865RSPT..155..459M} could not resolve the problem of divergences in the self-energy of a point charge. Therefore, Born-Infeld electrodynamics \cite{1934RSPSA.144..425B} was developed to address the issue. The theory introduces nonlinearity while preserving the fundamental properties of Maxwell's linear electrodynamics. It limits the maximum field strength that can be reached, preventing the theory from becoming unphysical. However, its quantization could not be achieved. A few year later, Heisenberg and Euler \cite{1936ZPhy...98..714H} proposed a theory that is a nonlinear extension of quantum electrodynamics that describes the behavior of the electromagnetic field in the presence of strong electric and magnetic fields. It takes into account the effects of virtual electron-positron pairs that arise due to the strong fields. However, this quantum electrodynamic theory, in general, does not guarantee the finiteness of the electric field at the origin \cite{2020IJMPA..3550211B}. Therefore, to determine a quantum electrodynamic theory without a singularity at the origin remained an open problem.

In the early 1940s, Bopp-Podolsky electrodynamics was constructed separately by Bopp \cite{Boppaaa} and Podolsky \cite{PhysRev.62.68aaa} that extends Maxwell’s linear electromagnetic theory. In Bopp-Podolsky electrodynamics, the Einstein field equations have terms up to fourth order arising from the Lagrangian comprising a new second order derivative of the gauge field. As we know that the ghost instabilities may arise in such higher order theories. However, Kaparulin et al. \cite{2014EPJC...74.3072K} developed a concept known as Lagrange anchor that is useful in avoiding instabilities in the Bopp-Podolsky electrodynamics. Cuzinatto et al. \cite{2007AnPhy.322.1211C} showed that the Bopp-Podolsky theory is a unique theory in $U(1)$ gauge group that is second order and linear generalization of Maxwell's theory. Furthermore, in the low energy limit, the Bopp-Podolsky term residing in the Lagrangian can be considered as an effective term arising due to quantum corrections to the photon action \cite{2019NuPhB.94414634B}. The coupling of Bopp-Podolsky electrodynamics with General Reativity (GR) is studied by Cuzinatto et al. \cite{2018EPJC...78...43Caaa}. They studied the spherically symmetric black holes (BHs) in Bopp-Podolsky electrodynamics and the no-hair theorem. The authors investigated the spacetime geometry without considering the analytic expression for geometry. It is also found that an extra parameter $b$ appears in the metric of the BHs, violating the no-hair theorem. Recently, Frizo et al. \cite{2023AnPhy.45769411Faaa} followed this approach for coupling of GR with the Bopp-Podolsky electrodynamics and attempted to derive a regular BH solution in spherical symmetry. They kept the mathematical scheme for both unknown metric functions and the electric field up to the linear approximation. However, they ended up with a wormhole solution of the Morris-Thorne type that admits an extra parameter $b$ in addition to mass $M$ and charge $Q$.

The goal of this study is to generate a rotating wormhole metric, whose static counterpart is given in Ref. \cite{2023AnPhy.45769411Faaa}, without resorting to the well-known Newman-Janis algorithm \cite{1965JMP.....6..915Naaa}. Instead of it, we employ the general algorithm provided in \cite{2014EPJC...74.2865Aaaa,2014PhRvD..90f4041Aaaa,2016EPJC...76....7Aaaa}. It is worth mentioning that, this procedure provides non-trivial rotating wormholes only if the static background has not a trivial red-shift function. Of course, the solution given in \cite{2023AnPhy.45769411Faaa} has a highly non-trivial lapse function satisfying all the wormhole requirements. To further validate the model, we explore the shadow cast and the deflection of light in the strong field regime for the obtained rotating wormhole. The wormhole spacetime we investigate is characterized by the ability to separate variables in the equations governing null geodesics. This feature simplifies the process of analyzing how light travels in this spacetime. As a result, we can analytically compute the boundary of the shadow, which is essentially the image created by the lensing effect of unstable spherical photon orbits around the wormhole. This approach provides a detailed understanding of the shadow's shape and structure, which is crucial for interpreting observational data and theoretical predictions. The unstable null orbits are categorized into two separate classes known as inner and outer spherical orbits. The inner unstable orbits are situated at the wormhole's throat, whereas, the outer unstable orbits are found at a greater radial distance from the throat. The formulation for the shadow curves corresponding to the outer null orbits was established in \cite{2013PhRvD..88l4019N}. Nevertheless, the analysis of the shadow presented in that study was not fully detailed. Later, h \cite{PhysRevD.98.024044} investigated the shadows for the inner unstable orbits. However, these shadow curves are flawed due to an inaccurate estimation of the mass of the wormhole, resulting in both qualitative and scale discrepancies from realistic representations. This problem was also addressed in Ref. \cite{2018EPJC...78..544G}, the existing results were summarized, and the structure of the shadow curves was comprehended. 

To capture the main features about the optical properties of this particular model sourced by this unusual electrodynamics. We analyze in detail the gravitational lensing effect in the so-called strong regime. In this concern, it should be taken into account that for such an analysis for a wormhole structure is quite different from that for a BH. This is because of the fact that for wormholes their throat plays a pivotal role \cite{h_2019aaa}. Here, as a first approach in this subject, we consider the case where the observer and source are on apposite side of the wormhole throat. Through Bozza's procedure \cite{Bozza:2002afaaa,Bozza:2002zjaaa,Bozza:2008evaaa}, we explore in detail the behavior of the deflection angle, Einstein rings and lensing observables, that is, the angular position of the asymptotic relativistic images $\left(\theta_\infty\right)$, angular separation between the outermost and asymptotic relativistic images $(s)$ and relative magnification of the outermost relativistic image with other relativistic images $(rmag)$.

The article is organized as follows: In Sect. \ref{sec2}, a short revision of the original wormhole solution and the Bopp-Podolsky electrodynamics is provided. The Sect. \ref{sec3} provides the rotating model, Sects. \ref{section4} and \ref{sec5} are devoted to the analysis of the optical properties of the wormhole spacetime and finally Sect. \ref{sec6} concludes the work.

Throughout the article we use units where $c=G=1$ and the mostly negative signature $(+;-;-;-)$.

\section{Bopp-Podolsky Electrodynamics and the Static Wormhole}\label{sec2}
We begin with a brief overview of the Einstein-Hilbert-Podolsky action describing the coupling of gravity and Bopp-Podolsky electrodynamics \cite{2023AnPhy.45769411Faaa}. Mathematically, it is given as
\begin{eqnarray}
\mathcal{S}=\frac{1}{4\pi}\int d^4x\sqrt{-g}\left(-\frac{R}{4}+\mathcal{L}_m\right), \label{1}
\end{eqnarray}
where, $g$ is the determinant of the metric tensor, $R$ is the Ricci scalar, and $\mathcal{L}_m$ is the matter Lagrangian for the Bopp-Podolsky electrodynamics field given as
\begin{eqnarray}
\mathcal{L}_m=-\frac{1}{4}F^{\mu\nu}F_{\mu\nu}+\frac{1}{2}\left(a^2+2b^2\right)\nabla_\nu F^{\mu\nu}F_{\gamma}F^{\gamma}_{\mu}+b^2\left(R_{\sigma\nu}F^{\sigma\mu}F_{\mu}^{\nu}+R_{\mu\sigma\nu\gamma}F^{\sigma\gamma}F^{\mu\nu}\right), \label{2}
\end{eqnarray}
such that $F_{\mu\nu}=\nabla_\mu A_\nu-\nabla_\nu A_\mu$ is the Maxwell's field tensor with $A_\mu$ and $\nabla_\mu$ being the four-potential and covariant derivative, respectively. Here, $R_{\mu\nu}$ and $R_{\mu\nu\sigma\gamma}$ are Ricci and Riemann tensors, respectively with $a$ and $b$ as the coupling constants. Besides the Maxwell's term in the matter Lagrangian for the Bopp-Podolsky electrodynamics, the other two terms are invariant and independent \cite{2018EPJC...78...43Caaa}. Therefore, the Lagrangian obeys four properties, that is, it becomes invariant under Lorentz transformations, becomes gauge invariant under the symmetry group $U(1)$, is quadratic in the gauge field and its derivatives, and is dependent on gauge field and its derivatives of order one and two. By varying the action (\ref{1}) with respect to the metric tensor $g_{\mu\nu}$, one gets the following Einstein field equations:
\begin{eqnarray}
R_{\mu\nu}-\frac{1}{2}g_{\mu\nu}R=8\pi T_{\mu\nu}=8\pi\left(T^M_{\mu\nu}+T^a_{\mu\nu}+T^b_{\mu\nu}\right), \label{3}
\end{eqnarray}
where, the components of energy-momentum tensor $T_{\mu\nu}$ are
\begin{eqnarray}
T^M_{\mu\nu}&=&\frac{1}{16\pi}\left[4F_{\mu\sigma}F^\sigma_\nu+g_{\mu\nu}F^{\alpha\beta}F_{\alpha\beta}\right], \label{4}\\
T^a_{\mu\nu}&=&\frac{a^2}{8\pi}\left[2g_{\mu\nu}F^\gamma_\beta\nabla_\gamma K^\beta+g_{\mu\nu}K^\beta K_\beta+4F_{(\mu}^\alpha\nabla_{\nu)} K_\alpha-4F_{(\mu}^\alpha\nabla_\alpha K_{\nu)}-2K_\mu K_\nu\right], \label{5}\\
T^b_{\mu\nu}&=&\frac{b^2}{8\pi}\left[-g_{\mu\nu}\nabla^\beta F^{\alpha\gamma}\nabla_\beta F_{\alpha\gamma}+4F^\gamma_{(\mu}\nabla^\beta\nabla_\beta F_{\nu)\gamma}+4F_{\gamma(\mu}\nabla_\beta\nabla_{\nu)}F^{\beta\gamma}-4\nabla_\beta\left(F_\gamma^\beta\nabla_{(\mu}F_{\nu)}^\gamma\right)\right], \label{6}
\end{eqnarray}
where, the notation $(...)$ represents the symmetrization with respect to the indices existing inside the brackets. Under the variation of the action (\ref{1}) with respect to the four potential $A_\mu$, one can obtain the Bopp-Podolsky equations for curved spacetime given as
\begin{eqnarray}
\nabla_\nu\left[F^{\mu\nu}-\left(a^2+2b^2\right)\left(\nabla^\mu K^\nu-\nabla^\nu K^\mu\right)+2b^2\left(F^{\mu\sigma}R_\sigma^\nu-F^{\nu\sigma}R_\sigma^\mu+2R^{\mu\nu}_{\sigma\beta}F^{\beta\sigma}\right)\right]=0, \label{7}
\end{eqnarray}
where, $K^\mu=\nabla_\nu F^{\mu\nu}$. The energy-momentum tensor is conserved, that is,
\begin{eqnarray}
\nabla_\nu T^{\mu\nu}=\nabla_\nu\left(T^M_{\mu\nu}+T^a_{\mu\nu}+T^b_{\mu\nu}\right)=0 \label{8}
\end{eqnarray}
by utilizing the constraint given in the Eq. (\ref{7}). It must be noted that any valid solution in the context  of Bopp-Podolsky framework must satisfy field equations (\ref{3}) or (\ref{7}). Despite of not deriving the exact form of the spacetime metric, the authors of Ref. \cite{2018EPJC...78...43Caaa} incorporated the Bekenstein method and showed that the exterior of a spherically symmetric BH must be Reissner-Nordstr\"{o}m solution when $b=0$. This satisfied the no-hair conjecture. The authors also proposed that hairy BH solutions can be obtained only if $b\neq0$ and $\frac{dg_{00}}{dr}\geq0$, where, $r$ denotes the radial coordinate. However, Frizo et al. \cite{2023AnPhy.45769411Faaa} showed that by incorporating a perturbative approach, $b\neq0$ generates a wormhole solution instead of a BH one. They considered the static and spherically symmetric spacetime given as
\begin{eqnarray}
ds^2=f(r)dt^2-\frac{dr^2}{g(r)}-r^2d\theta^2-r^2\sin^2\theta d\phi^2, \label{9}
\end{eqnarray}
where, $f(r)\neq g(r)$ in general. The values of these functions can be obtained by considering the first order approximations,
\begin{eqnarray}
f(r)=f_0(r)+\epsilon f_1(r), \label{10}\\
g(r)=g_0(r)+\epsilon g_1(r), \label{11}\\
E(r)=E_0(r)+\epsilon E_1(r), \label{12}
\end{eqnarray}
to keep the perturbation up to linear order. Here, $E(r)$ is the electric field and $\epsilon=\epsilon(a,b)$ is a small parameter. Since, the required geometry must deviate from Reissner-Nordstr\"{o}m metric by a small amount, therefore, it can be assumed that
\begin{eqnarray}
f_0(r)=g_0(r)=1-\frac{2M}{r}+\frac{Q^2}{r^2}, \label{13}\\
E_0(r)=\frac{Q}{r^2}, \label{14}
\end{eqnarray}
whereas, the field strength is considered as
\begin{eqnarray}
F_{\mu\nu}=E(r)\left(\delta^1_\mu\delta^0_\nu-\delta^0_\mu\delta^1_\nu\right), \label{15}
\end{eqnarray}
where, $\delta^\alpha_\beta$ is the Kronecker delta function. By considering all of these assumptions and solving the field equations, the metric (\ref{9}) then takes the required approximate spacetime form as
\begin{eqnarray}
ds^2&=&\left(1-\frac{2M}{r}+\frac{Q^2}{r^2}+\frac{2b^2Q^2}{r^4}-\frac{6Mb^2Q^2}{r^5}+\frac{18b^2Q^4}{5r^6}\right)dt^2\nonumber\\&&-\left(1-\frac{2M}{r}+\frac{Q^2}{r^2}-\frac{4b^2Q^2}{r^4}+\frac{6Mb^2Q^2}{r^5}-\frac{12b^2Q^4}{5r^6}\right)^{-1}dr^2-r^2d\theta^2-r^2\sin^2\theta d\phi^2, \label{16}
\end{eqnarray}
such that $f(r)\neq g(r)$. It can be noted that by setting $b=0$, one can recover Reissner-Nordstr\"{o}m metric. Moreover, by considering $Q=0$, the metric behaves as Schwarzschild BH. The electric field becomes
\begin{eqnarray}
E(r)=\frac{Q}{r^2}\left(1-\frac{8Mb^2}{r^3}+\frac{11b^2Q^2}{r^4}\right), \label{17}
\end{eqnarray}
such that under the limit $b\to0$, one gets $E(r)\to E_0(r)$. We can see that at large distances, the effect of $b$ on electric field is very small. Therefore, the asymptotic behavior of the electric field is also the usual field $E_0(r)$. It is worth mentioning that the approximate spacetime metric (\ref{16}) is asymptotically flat, that is, $f(r)\to1$ and $g(r)\to1$ as $r\to\infty$.

\begin{figure}[t]
\centering
\subfigure{\includegraphics[width=0.4\textwidth]{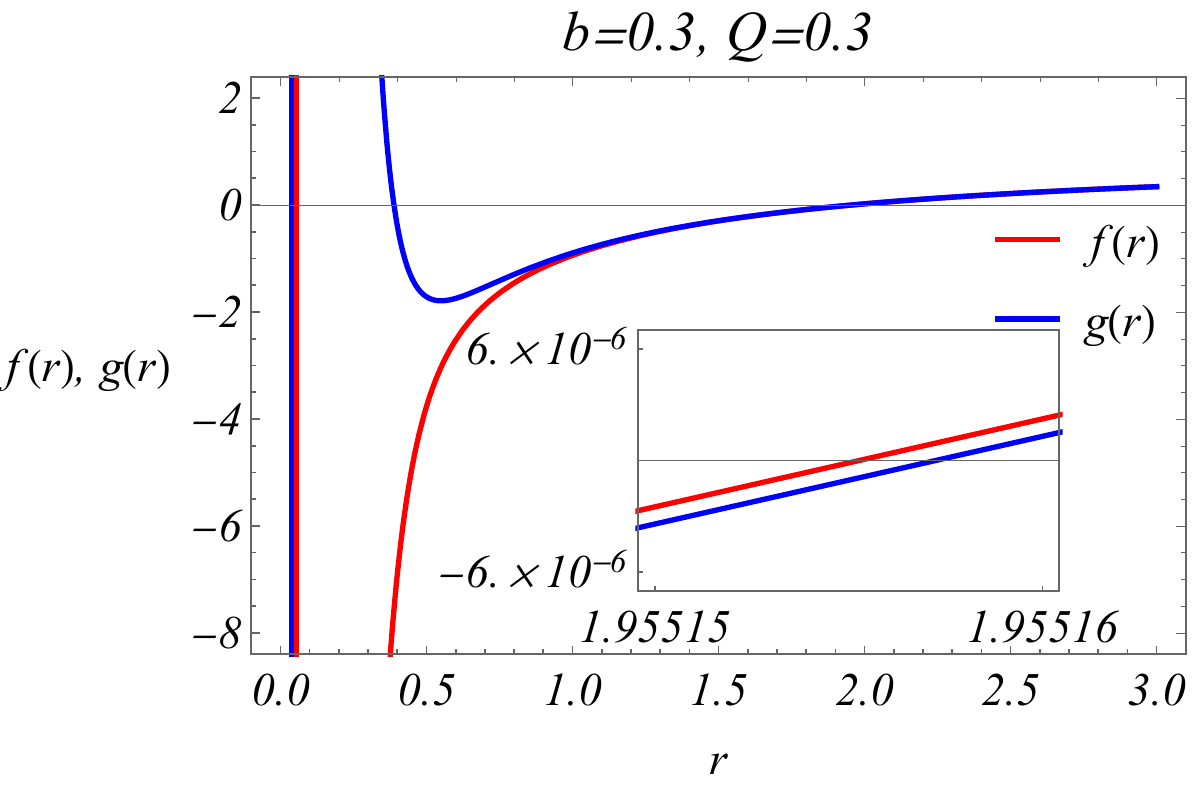}}~~~~~
\subfigure{\includegraphics[width=0.4\textwidth]{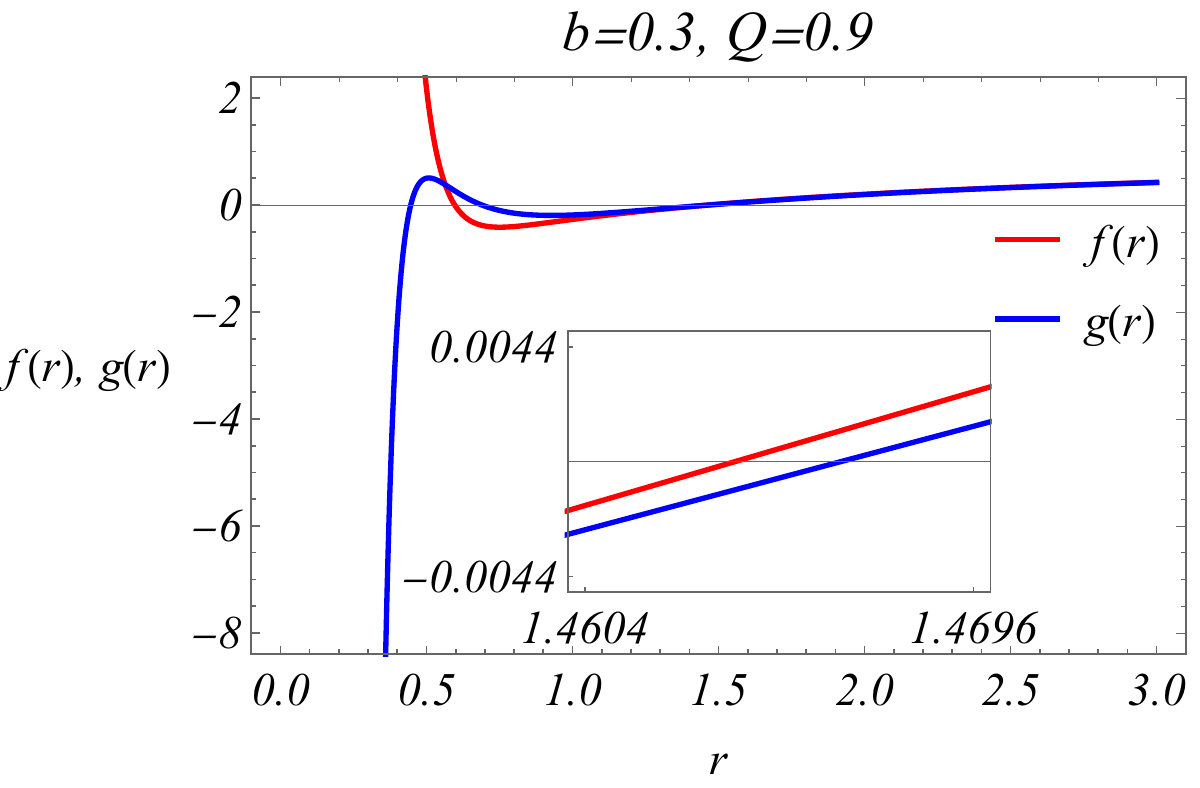}}\\
\subfigure{\includegraphics[width=0.4\textwidth]{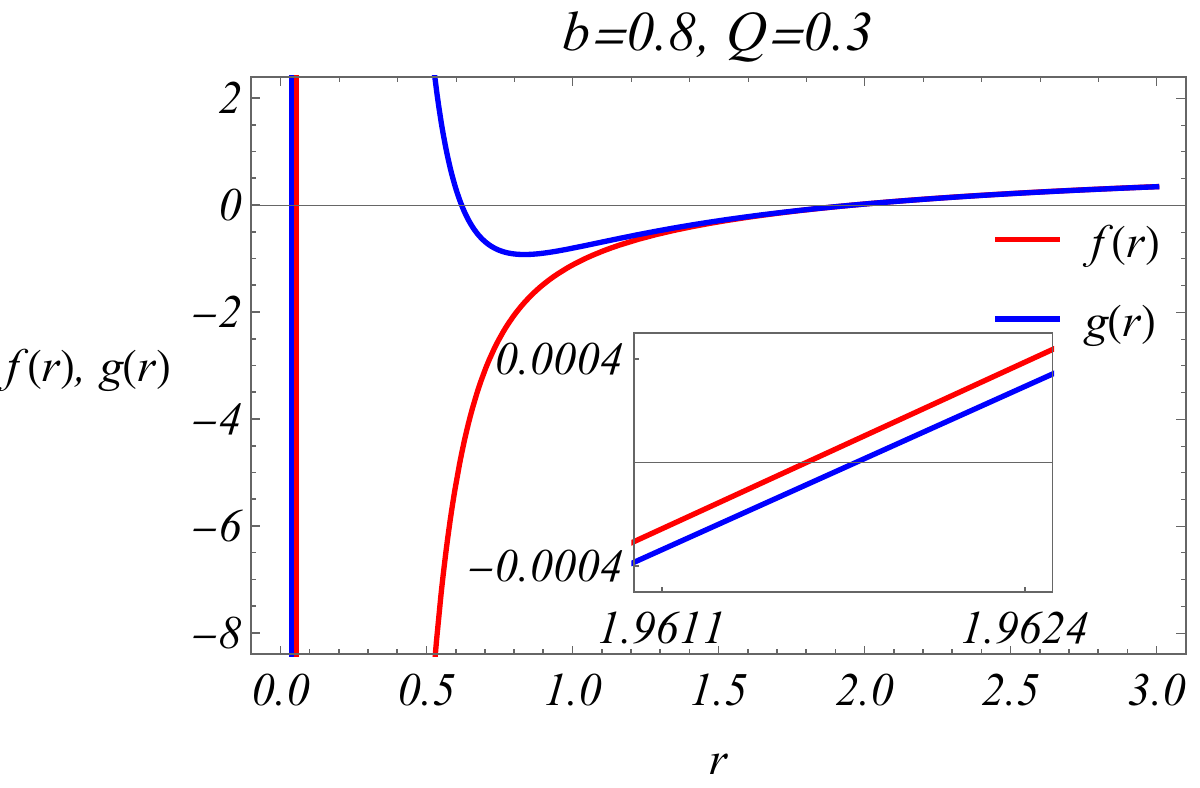}}~~~~~
\subfigure{\includegraphics[width=0.4\textwidth]{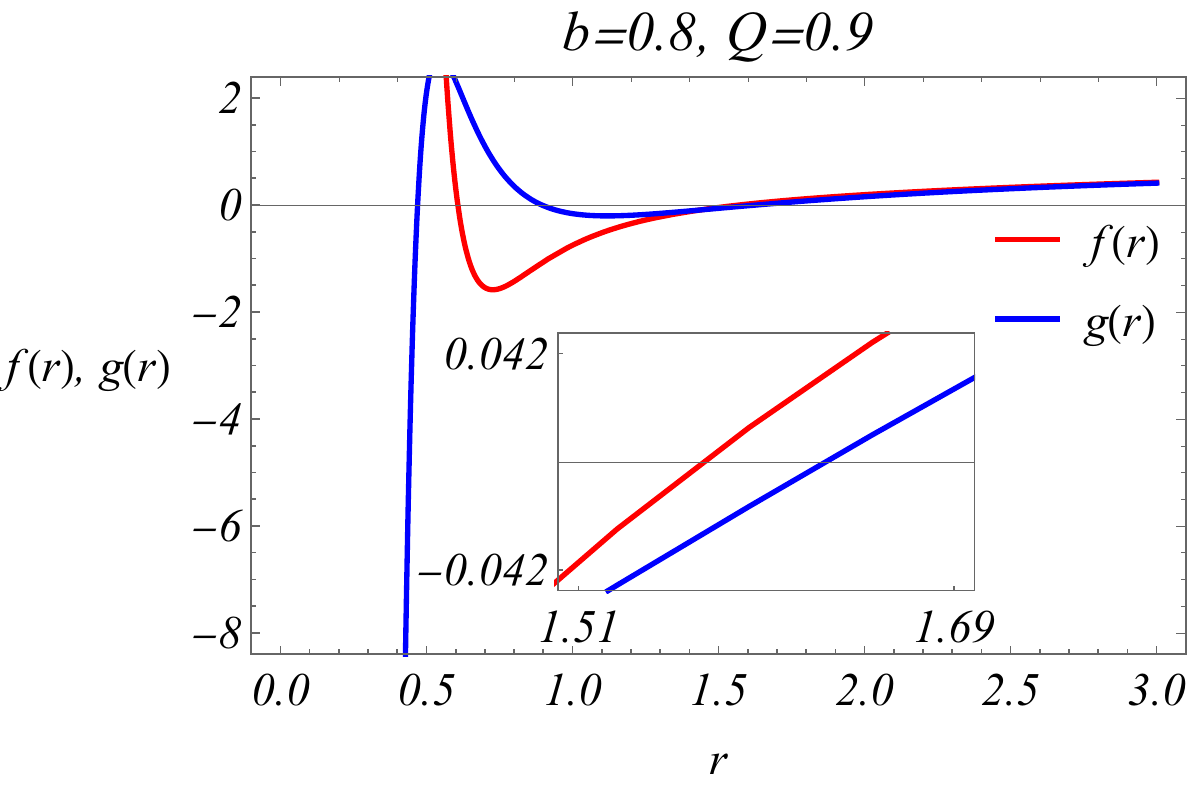}}
\caption{The behavior of $f(r)$ and $g(r)$ vs $r$ for different values of $b$ and $Q$ with $M=1$. \label{MF}}
\end{figure}

As mentioned above that Cuzinatto et al. \cite{2018EPJC...78...43Caaa} without deriving an explicit final form of the metric showed that under the conditions $b\neq0$ and $\frac{dg_{00}}{dr}\geq0$, the solution describes a hairy BH in Bopp-Podolsky electrodynamics. In this context, if $f(r)=0$ and $g(r)=0$ have same roots, then the roots of $g(r)=0$ are horizons of the spacetime. In particular, for $M>Q$, the equation $g(r)=0$ admits three real and positive roots out of which, the largest root is the event horizon. We know that $f(r)\neq g(r)$ and the metric signature is Lorentzian. However, for the values $r\leq r_+$, the metric signature is non-Lorentzian because the largest roots of $f(r)$ and $g(r)$ are not equal. This can be seen in Fig. \ref{MF}. For various combinations of parametric values for $b$ and $Q$, we have plotted the metric functions $f(r)$ and $g(r)$ with respect to $r$. The inset plots show the difference in roots of the functions $f(r)$ and $g(r)$. In all plots, we can see that there are three roots of $g(r)$. Moreover, the behavior of the metric functions $f(r)$ and $g(r)$ with respect to $b$ and $Q$ is very much same near the largest root of $g(r)$ in all plots. Furthermore, the largest root of $g(r)$ decreases with increase in $Q$, whereas, no significant impact of $b$ is observed on the size of the largest root of $g(r)$. The metric functions $f(r)$ and $g(r)$ are not much different from each other near the largest root of $g(r)$ under the prescribed limit of $b$. In this context, we can regard $b$ as a less sensitive parameter unlike $Q$. However, Frizo et al. \cite{2023AnPhy.45769411Faaa} mentioned that the surface $r=r_+$ would not be a null surface for which the largest root of $g(r)$ cannot be regarded as an event horizon. Therefore, $ds^2|_{r_+}\neq0$ for constant values of $r$, $\theta$ and $\phi$. To overcome this problem, the approximate metric (\ref{16}) is regarded as a wormhole solution instead of a BH solution. Considering this, the radial coordinate $r$ is valid outside the surface $r_+$ and hence, $r_+=r_{thr}=r_0$ is interpreted as wormhole throat point. Therefore, corresponding to each combination of values of parameters $b$ and $Q$, the largest root of $g_{rr}=g(r)=0$ generates a wormhole throat at $r_0$ instead of an event horizon. Consequently, the functions $f(r)$ and $g(r)$ are positive-definite for $r>r_0$. Moreover, the spacetime metric is Lorentzian and the chart $(t,r,\theta,\phi)$ is valid in the interval $r\in(r_0,\infty)$.

\section{Morris-Thorne Form and the Rotation}\label{sec3}
The Morris-Thorne wormhole is describe by the metric \cite{1988AmJPh..56..395M}
\begin{eqnarray}
ds^2=A(r)dt^2-\frac{dr^2}{1-\frac{b(r)}{r}}-r^2d\theta^2-r^2\sin^2\theta d\phi^2, \label{18}
\end{eqnarray}
where, $A(r)$ and $b(r)$ define the lapse and the shape functions, respectively. The throat point $r_0$ of the wormhole is determined by the roots of the equation $1-\frac{b(r)}{r}=0$ and is defined by the equation $r=r_0=b(r_0)$. Interestingly, the wormhole spacetime (\ref{16}) can be converted into the form (\ref{18}) by appropriately choosing the functions $A(r)$ and $b(r)$ in terms of $f(r)$ and $g(r)$, respectively. That is, if we consider
\begin{eqnarray}
A(r)&=&f(r), \label{19}\\
b(r)&=&r\left(1-g(r)\right)=2M-\frac{Q^2}{r}+\frac{4b^2Q^2}{r^3}-\frac{6Mb^2Q^2}{r^4}+\frac{12b^2Q^4}{5r^5}, \label{20}
\end{eqnarray}
then the metric (\ref{16}) can be interpreted as Morris-Thorne wormhole (\ref{18}). The functions $A(r)$ and $b(r)$, and hence $f(r)$ and $g(r)$ satisfy the following properties:
\begin{eqnarray}
\lim\limits_{r\to\infty}A(r)=\lim\limits_{r\to\infty}f(r)&=&1, \label{21}\\
\lim\limits_{r\to\infty}\frac{b(r)}{r}=\lim\limits_{r\to\infty}\left(1-g(r)\right)&=&0, \label{22}\\
\lim\limits_{r\to\infty}b(r)=\lim\limits_{r\to\infty}r\left(1-g(r)\right)&=&2M, \label{23}\\
\partial_rb(r_0)\leq1 \quad \text{or} \quad \partial_rg(r_0)&\geq&0, \label{24}\\
b(r)<r \quad \text{or} \quad g(r)>0 \quad \text{if} \quad r&>&r_0, \label{25}\\
r\partial_rb(r)<b(r) \quad \text{or} \quad \partial_rg(r)>0 \quad \text{if} \quad r&\geq&r_0. \label{26}
\end{eqnarray}

It is worth mentioning that the metric (\ref{16}) thus satisfies all of the conditions given in the Eqs. (\ref{21})-(\ref{26}) for certain parametric values. We performed a careful analysis for $M=1$, $b_i\in\left\{0,0.02,0.04,0.06,...,1\right\}$ and $Q_i\in\left\{0,0.02,0.04,0.06,...,1\right\}$ for each combination $(b_i,Q_i)$. We found that the conditions (\ref{24})-(\ref{26}) are not satisfied for the combinations $(0,Q_i)$ (of course, this corresponds to the Kerr-Newman BH), $(b_i,0)$, $(0.02,0.02)$, $(0.02,0.04)$ and $(0.04,0.02)$. This ultimately provides us limits for the parametric values. Moreover, the valid parametric values also satisfy the property that the largest root of $g(r)$ is greater than the largest root of $f(r)$. This fact is also obvious from the plots in Fig. \ref{MF}. For smaller values of $b$ and $Q$, a very small difference in roots of $f(r)$ and $g(r)$ is observed. Whereas, for larger values of the parameters, the difference in these roots becomes more visible.

Azreg-A\"{i}nou \cite{2016EPJC...76....7Aaaa} derived the exact rotating counterpart for the static Morris-Thorne wormhole (\ref{18}) by incorporating the modified Newman-Janis algorithm constructed by himself \cite{2014PhRvD..90f4041Aaaa,2014EPJC...74.2865Aaaa}. He developed this algorithm by a slight modification in the Newman-Janis algorithm \cite{1965JMP.....6..915Naaa} in order to resolve the issue of its applicability in modified theories of gravity. In mid 60s, the Kerr and Kerr-Newman BH metrics were derived from the Schwarzschild and Reissner-Nordstr\"{o}m metrics, respectively, by using Newman-Janis algorithm. Though, it was mainly designed to convert the static BH spacetimes of GR into their rotating counterparts. However, some problems occurred in using this algorithm outside of GR. As we know that the Schwarzschild and Kerr BHs are vacuum solutions, whereas the Reissner-Nordstr\"{o}m and Kerr-Newman BHs are sourced by the electric field. In this context, we can propose that the NJA may not alter the source when applied to BHs in GR. However, Hansen and Yunes \cite{2013PhRvD..88j4020H} found that it is not true in the case of BHs in theories of modified gravity because they encountered some extra sources in the field equations. The complexification of the radial coordinate is not taken into account in the modified method. Instead, some arbitrary functions are defined that are determined as the algorithm proceeds. Therefore, it makes the modified method easier to deal with and is useful in determining numerous rotating BH as well as wormhole solutions in GR and beyond. The rotating counterpart of the static metric (\ref{18}) and hence the metric (\ref{16}) as derived in Ref. \cite{2016EPJC...76....7Aaaa} is given in Kerr-like form as
\begin{eqnarray}
ds^2&=&\left(1-\frac{2\zeta(r)}{\rho^2}\right)dt^2-\frac{\rho^2A(r)}{\Delta(r)\left(1-\frac{b(r)}{r}\right)}dr^2-\rho^2d\theta^2+\frac{4a\zeta(r)\sin^2\theta}{\rho^2}dtd\phi \nonumber\\&&-\frac{\sin^2\theta}{\rho^2}\left(\left(r^2+a^2\right)^2-\Delta(r)a^2\sin^2\theta\right)d\phi^2, \label{27}
\end{eqnarray}
where,
\begin{eqnarray}
\Delta(r)&=&a^2+r^2A(r)=r^2+a^2-2Mr+Q^2+\frac{2b^2Q^2}{r^2}-\frac{6Mb^2Q^2}{r^3}+\frac{18b^2Q^4}{5r^4}, \label{28}\\
\rho^2&=&r^2+a^2\cos^2\theta, \label{29}\\
\zeta(r)&=&\frac{r^2\left(1-A(r)\right)}{2}, \label{30}
\end{eqnarray}
with $a$ defining the spin of wormhole. The rotating metric (\ref{27}) can be written in the following Kerr-like form as well:
\begin{eqnarray}
ds^2&=&\left(\frac{\Delta(r)-a^2\sin^2\theta}{\rho^2}\right)dt^2-\frac{\rho^2A(r)}{\Delta(r)\left(1-\frac{b(r)}{r}\right)}dr^2-\rho^2d\theta^2+\frac{2a\sin^2\theta}{\rho^2}\left(r^2+a^2-\Delta(r)\right)dtd\phi \nonumber\\
&&-\frac{\sin^2\theta}{\rho^2}\left(\left(r^2+a^2\right)^2-\Delta(r)a^2\sin^2\theta\right)d\phi^2, \label{31}
\end{eqnarray}
where, $\Delta(r)$ and $\rho^2$ are defined by the Eqs. (\ref{28}) and (\ref{29}), respectively. Since, the metric (\ref{31}) is the exact rotating counterpart of the static metric (\ref{18}). Therefore, it is also the rotating counterpart of the static metric (\ref{16}).

It must be noted that under the limit $a\to0$, the rotating metric (\ref{31}) reduces to the static metric (\ref{16}). When we consider $b=0$ in the rotating metric (\ref{31}), we get the Kerr-Newman BH in the same way as we get Reissner-Nordstr\"{o}m BH in the static case. Furthermore, under the limit $Q=0$, the metric (\ref{31}) reduces to Kerr BH as we get Schwarzschild BH in the static case. Since, the metric (\ref{31}) exhibits the ratio $g_{\theta\theta}/g_{rr}$ to be independent of the coordinate $\theta$. Therefore, the metric (\ref{31}) can be written in the form
\begin{eqnarray}
ds^2=\frac{\Delta(r)}{\rho^2}\left(dt-a\sin^2\theta d\phi\right)^2-\frac{\rho^2A(r)}{\Delta(r)\left(1-\frac{b(r)}{r}\right)}dr^2-\rho^2d\theta^2-\frac{\sin^2\theta}{\rho^2}\left[\left(r^2+a^2\right)d\phi-adt\right]^2. \label{32}
\end{eqnarray}

Following the form in the metric (\ref{32}), the symmetry properties of the stationary and axisymmetric spacetime are satisfied. Moreover, the metric (\ref{31}) is independent of the coordinates $t$ and $\phi$. Therefore, the metric possesses two Killing vector fields, $\chi^\mu_{(t)}=\left(\partial_t\right)^\mu$ and $\chi^\mu_{(\phi)}=\left(\partial_\phi\right)^\mu$ known as the time translational and azimuthal Killing vectors, respectively. If
\begin{eqnarray}
\frac{\Delta(r)\left(1-\frac{b(r)}{r}\right)}{\rho^2A(r)}=1-\frac{b_r(r,\theta)}{r}, \label{33}
\end{eqnarray}
then the metric (\ref{31}) is reduced to the Morris-Thorne form and the function $b_r(r,\theta)$ will define the shape function of the rotating wormhole. Under the analogy with the static wormholes, the throat is defined by the surface $b_r(r_0,\theta_0)=r_0$. Since the equation $b_r(r_0,\theta_0)=r_0$ implies that the function $1-b_r(r,\theta)/r=0$ at the throat. Therefore, following the Eq. (\ref{33}), we get $b(r_0)=r_0$ because in general, $\Delta(r)\neq0$ at the throat. Therefore, the throat is formed at the radial distance $r=r_0$ which is same as the throat point of the static wormhole (\ref{16}), that is, the spin has no effect on the location of the throat. In the static case, the functions $A(r)$ and $b(r)$ satisfied some properties given in the Eqs. (\ref{21})-(\ref{26}). Likewise, the rotating metric (\ref{31}) also satisfies the corresponding conditions. That is,
\begin{eqnarray}
\lim\limits_{r\to\infty}\left(\frac{\Delta(r)-a^2\sin^2\theta}{\rho^2}\right)&=&1, \label{34}\\
\lim\limits_{r\to\infty}\frac{b_r(r,\theta)}{r}=1-\frac{\Delta(r)\left(1-\frac{b(r)}{r}\right)}{\rho^2A(r)}&=&0, \label{35}\\
\lim\limits_{r\to\infty}b_r(r,\theta)=r\left(1-\frac{\Delta(r)\left(1-\frac{b(r)}{r}\right)}{\rho^2A(r)}\right)&=&2M. \label{36}
\end{eqnarray}
To check the validity of the flare-out and other remaining conditions, we take derivative of the Eq. (\ref{33}) with respect to $r$, then we get
\begin{eqnarray}
\frac{b_r(r,\theta)-r\partial_rb_r(r,\theta)}{r^2}=\frac{\Delta(r)}{\rho^2A(r)}\frac{b(r)-r\partial_rb(r)}{r^2}+\left(1-\frac{b(r)}{r}\right)\partial_r\left(\frac{\Delta(r)}{\rho^2A(r)}\right). \label{37}
\end{eqnarray}
Since $1-\frac{b(r)}{r}\to0$ as $r\to r_0$, so the second term on the right hand side in above Eq. (\ref{37}) vanishes and the factor $\frac{\Delta(r)}{\rho^2A(r)}>0$ everywhere on and outside the throat, then the flare-out condition in the Eq. (\ref{26}) implies that the corresponding flare-out condition for rotating wormhole (\ref{31}) also holds. That is,
\begin{eqnarray}
r\partial_rb_r(r,\theta)<b_r(r,\theta) \quad \text{if} \quad r\geq r_0. \label{38}
\end{eqnarray}
Now, if we consider $b_r(r_0,\theta_0)=r_0$ and $b(r_0)=r_0$ in the Eq. (\ref{37}), the second term on the right hand side in the Eq. (\ref{37}) vanishes and we get
\begin{eqnarray}
1-\partial_rb_r(r_0,\theta_0)=\frac{\Delta(r_0)}{\rho^2A(r_0)}\left(1-\partial_rb(r_0)\right) \label{39}
\end{eqnarray}
and from Eq. (\ref{24}), $\partial_rb(r_0)\leq1$ with the factor $\frac{\Delta(r_0)}{\rho^2A(r_0)}>0$ as mentioned above, we get
\begin{eqnarray}
1-\partial_rb_r(r_0,\theta_0)\geq0  \quad \text{or} \quad \partial_rb_r(r_0,\theta_0)\leq1. \label{40}
\end{eqnarray}
Lastly, from the condition $b(r)<r$ outside the throat in the static case, the left hand side of Eq. (\ref{37}) is greater than zero. This implies that the right hand side of Eq. (\ref{37}) is also greater than zero. Therefore, we get
\begin{eqnarray}
b_r(r,\theta)\leq r  \quad \text{if} \quad r>r_0. \label{41}
\end{eqnarray}
Since the properties satisfied in the Eqs. (\ref{38}), (\ref{40}) and (\ref{41}) are exactly analogous to those for the static case in the Eqs. (\ref{24})-(\ref{26}). Therefore, the rotating wormhole (\ref{31}) satisfies all of the properties corresponding to the Eqs. (\ref{21})-(\ref{26}) for the static case.

\section{Shadow}\label{section4}
Photons emerging out of a luminous source are trapped in a critical region around the throat of a wormhole. This region is made up of circular null trajectories that are either stable or unstable. Since, the shadow formation is due to the unstable null orbits, therefore, we are only interested in the unstable null circular trajectories. The circular photon orbits are described by the effective potential function and the local maximum exhibited by this function describes the unstable orbits. The light particles in the unstable orbits either disappear into the wormhole or scatter away from the gravitational field of the wormhole throat towards infinity. Hence, due to the absence of photons, a 2D dark silhouette is formed better known as the shadow. We aim to study the shadow of the rotating wormhole in the Bopp-Podolsky electrodynamics. For which we need to study the properties of null geodesics and effective potential. We develop the mathematical scheme for the analysis by employing the Hamiltonian and Hamilton-Jacobi formalisms. The system of equations are completely integrable if we determine four constants of motion as system comprises four coordinate variables. The Hamiltonian method generates only two constants known as the energy $E$ and angular momentum $L$ along $z$ direction. The mass of the particle can be considered as the third constant. However, the Hamilton-Jacobi formalism is helpful in obtaining the fourth constant that we call as Carter constant $\mathcal{Z}$ \cite{PhysRev.174.1559}. To proceed, it is important that the mixed terms in $r$ and $\theta$ are separated for which the Hamilton-Jacobi method is useful. The Hamiltonian describing the motion of photon is written as
\begin{eqnarray}
2\mathcal{H}=g^{\mu\nu}p_\mu p_\nu&=&\frac{\left(r^2+a^2\right)^2-a^2\Delta(r)\sin^2\theta}{\Delta(r)\rho^2}p_t^2+\frac{\Delta(r)\left(b(r)-r\right)}{rA(r)\rho^2}p_r^2-\frac{1}{\rho^2}p_\theta^2 \nonumber\\
&&+\frac{a^2\sin^2\theta-\Delta(r)}{\Delta(r)\rho^2\sin^2\theta}p_\phi^2+\frac{2a\left(r^2+a^2-\Delta(r)\right)}{\Delta(r)\rho^2}p_tp_\phi, \label{ham1}
\end{eqnarray}
where, $p_\mu$ is momentum of the particle. The Hamilton's equations
\begin{equation}
\dot{x}^\mu=\partial_\tau x^\mu=\partial_{p_\mu}\mathcal{H}, \qquad \dot{p}_\mu=\partial_\tau p_\mu=-\partial_{x^\mu}\mathcal{H} \label{hameq}
\end{equation}
are required to solve the Hamiltonian (\ref{ham1}) for the equations of motion. Note that the symbol $\tau$ is denotes an affine parameter. Therefore the velocity equations corresponding to $t$ and $\phi$ coordinates come out to be
\begin{eqnarray}
\rho^2\dot{t}&=&\frac{r^2+a^2}{\Delta(r)}\left(\left(r^2+a^2\right)E-aL\right)+a\left(L-aE\sin^2\theta\right), \label{tdot} \\
\rho^2\dot{\phi}&=&\frac{a}{\Delta(r)}\left(\left(r^2+a^2\right)E-aL\right)+\left(L\csc^2\theta-aE\right). \label{phidot}
\end{eqnarray}
The constants $E$ and $L$ have been obtained as the energy and angular momentum of the particle, respectively in above Eqs. (\ref{tdot}) and (\ref{phidot}). Whereas, we assume photon mass $m_p=0$ as the third constant such that the Hamiltonian $\mathcal{H}=0$. Therefore, the relation $p_\mu=\partial_{x^\mu}\mathcal{S_J}$ enables one to write the Hamilton-Jacobi equation in the following form
\begin{equation}
g^{\mu\nu}\partial_{x^\mu}\mathcal{S_J}\partial_{x^\nu}\mathcal{S_J}=0 \label{HJ}
\end{equation}
such that the Jacobi action
\begin{equation}
\mathcal{S_J}=-Et+\mathcal{T}_r(r)+\mathcal{T}_\theta(\theta)+L\phi \label{JAc}
\end{equation}
comprises the unknown functions $\mathcal{T}_r(r)$ and $\mathcal{T}_\theta(\theta)$ that will be determined later. Expanding the Hamilton-Jacobi equation by using the Jacobi action, we get
\begin{equation}
\frac{\Delta(r)\left(r-b(r)\right)}{rA(r)}\big(\partial_r\mathcal{T}_r(r)\big)^2-\frac{\left(\left(r^2+a^2\right)E-aL\right)^2}{\Delta(r)}+\left(L-aE\right)^2+\left(\partial_\theta\mathcal{T}_\theta(\theta)\right)^2+\left(L^2\csc^2\theta-a^2E^2\right)\cos^2\theta=0, \label{mixhj}
\end{equation}
which is clearly separable in the coordinates $r$ and $\theta$. Therefore, it is separated as follows:
\begin{eqnarray}
\frac{\Delta(r)\left(r-b(r)\right)}{rA(r)}\big(\partial_r\mathcal{T}_r(r)\big)^2-\frac{\left(\left(r^2+a^2\right)E-aL\right)^2}{\Delta(r)}+\left(L-aE\right)^2&=&-\mathcal{Z}, \label{sepr}\\
\left(\partial_\theta\mathcal{T}_\theta(\theta)\right)^2+\left(L^2\csc^2\theta-a^2E^2\right)\cos^2\theta&=&\mathcal{Z}. \label{sept}
\end{eqnarray}
On further simplification using the definitions of Hamilton's equations, we obtain the other two geodesic equations in $r$ and $\theta$ directions as
\begin{eqnarray}
\rho^2\dot{r}&=&\sqrt{\mathcal{R}(r)}, \label{rg}\\
\rho^2\dot{\theta}&=&\sqrt{\Theta(\theta)}, \label{tg}\\
\end{eqnarray}
where,
\begin{figure}[t]
\centering
\subfigure{\includegraphics[width=0.4\textwidth]{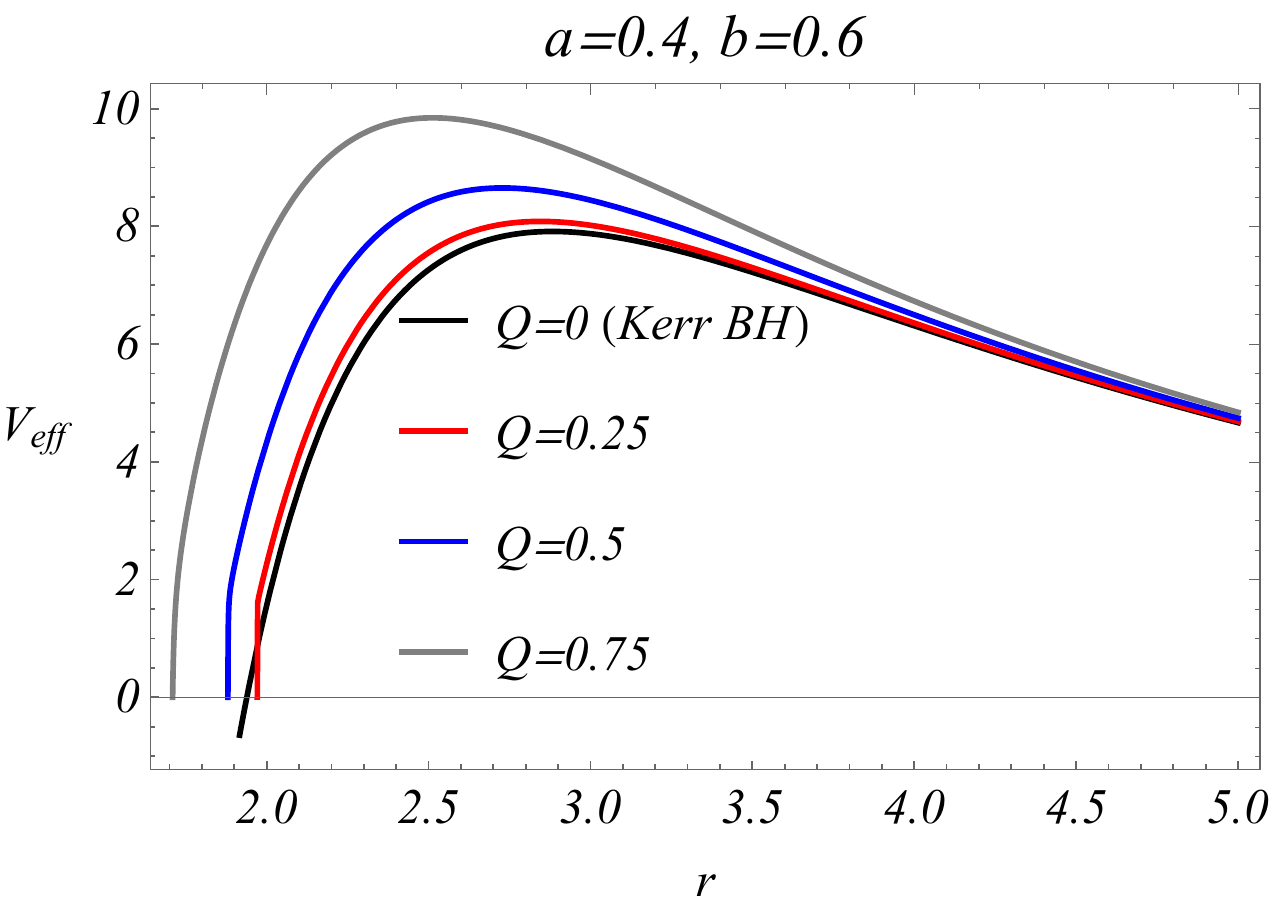}}~~~~~
\subfigure{\includegraphics[width=0.4\textwidth]{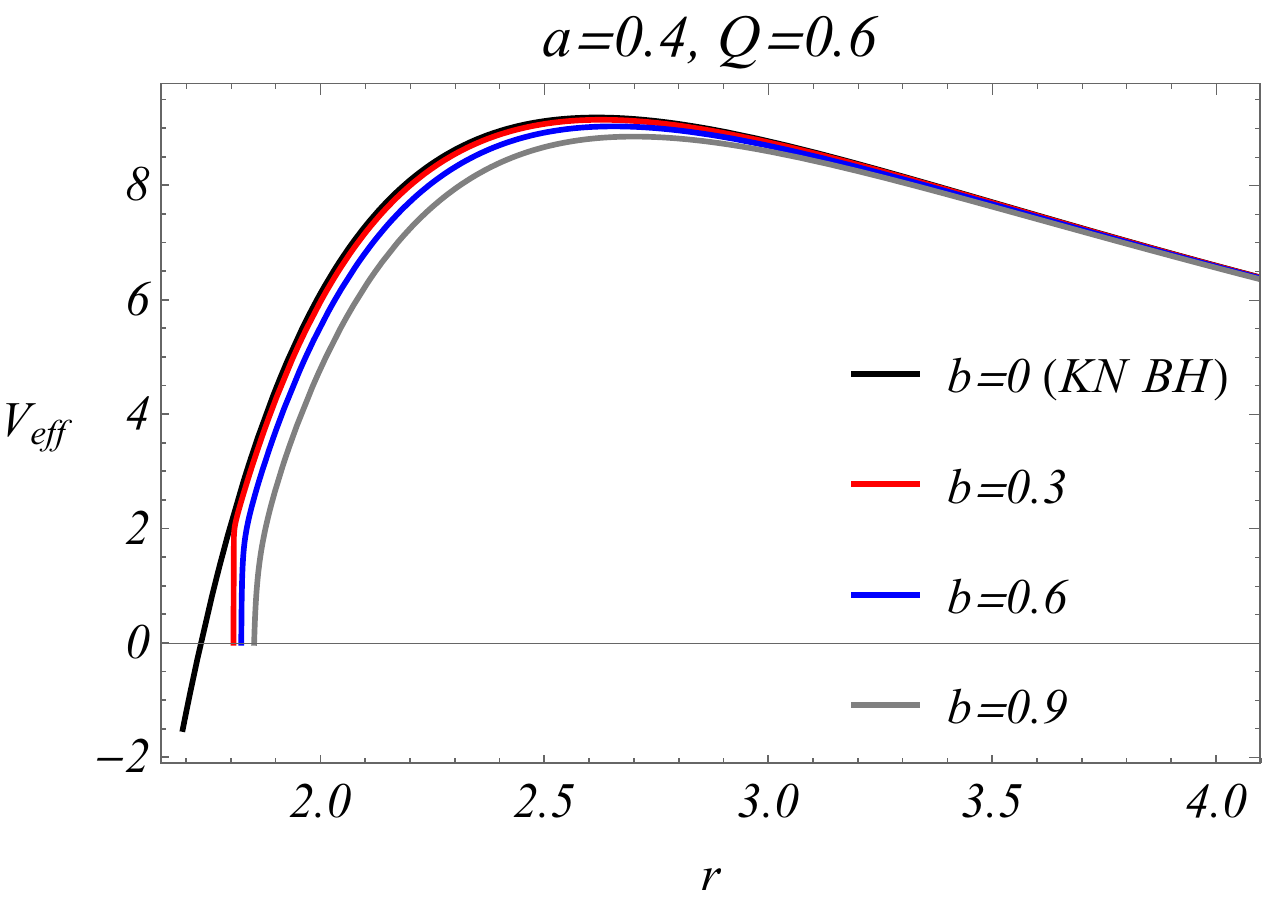}}\\
\subfigure{\includegraphics[width=0.4\textwidth]{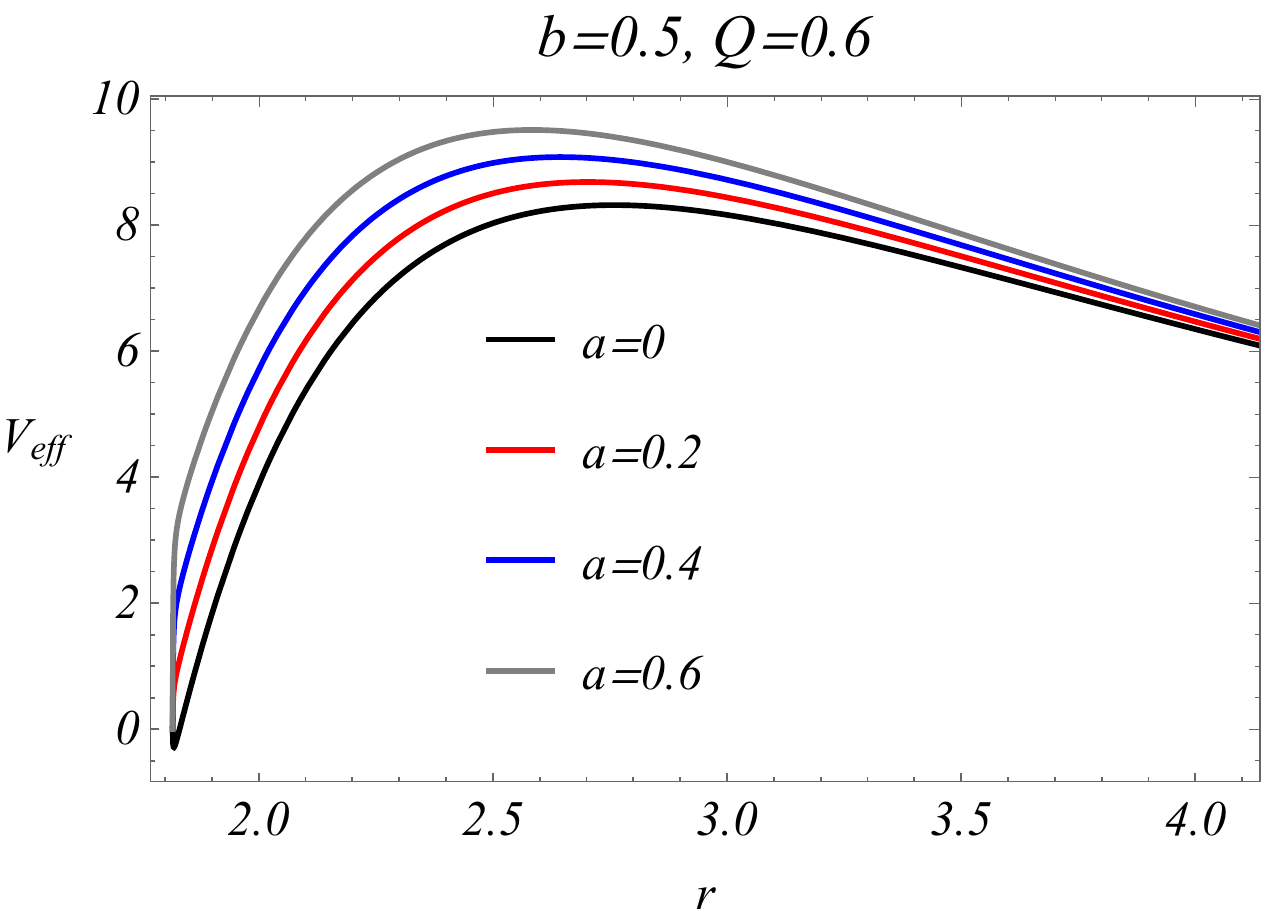}}
\caption{Effective potential curves showing the behavior of unstable circular null orbits for different values of $Q$, $b$ and $a$ corresponding to each curve in respective plots with $M=1$. \label{EP}}
\end{figure}
\begin{eqnarray}
\mathcal{R}(r)&=&\frac{r-b(r)}{rA(r)}\left[\left(\left(r^2+a^2\right)E-aL\right)^2-\Delta(r)\left(\mathcal{Z}+\left(L-aE\right)^2\right)\right], \label{Rfunc}\\
\Theta(\theta)&=&\mathcal{Z}-\left(L^2\csc^2\theta-a^2E^2\right)\cos^2\theta. \label{thfunc}
\end{eqnarray}
If we rename the constants of motion in terms of impact parameters as $u=\frac{L}{E}$ and $\Sigma=\frac{\mathcal{Z}}{E^2}$, then the functions $\mathcal{R}(r)$ and $\Theta(\theta)$ become
\begin{eqnarray}
\mathcal{R}(r)&=&\frac{r-b(r)}{rA(r)}\left[\left(r^2+a^2-au\right)^2-\Delta(r)\left(\Sigma+\left(u-a\right)^2\right)\right], \label{Rfuncn}\\
\Theta(\theta)&=&\Sigma-\left(u^2\csc^2\theta-a^2\right)\cos^2\theta. \label{thfuncn}
\end{eqnarray}
In order to analyze the effective potential, we consider the equatorial motion, that is, $\theta=\frac{\pi}{2}$ must be considered throughout the analysis. The radial geodesic equation is connected with the effective potential $V_{eff}$ in the form given as
\begin{eqnarray}
\frac{1}{2}\dot{r}^2+V_{eff}(r)=0, \label{veff}
\end{eqnarray}
where, $V_{eff}(r)$ is the effective potential of the moving particle (photon) which is explicitly given as
\begin{eqnarray}
V_{eff}(r)=-\frac{\mathcal{R}(r)}{2r^4}=\frac{b(r)-r}{2r^5A(r)}\left[\left(r^2+a^2-au\right)^2-\Delta(r)\left(\Sigma+\left(u-a\right)^2\right)\right]. \label{veff1}
\end{eqnarray}

The effective potential describes the behavior of the particle under motion in the photon region. It also describes the characteristics of the orbits in the photon region. Since, we are mainly interested in investigating the shape and size of the wormhole shadow that is solely a property of the light particles disappearing as a result of falling in the throat of the wormhole from the unstable circular photon orbits. Since the photon trajectories are circular orbits, therefore, the photons must satisfy the equation $r=\text{constant}$. Therefore, these circular light trajectories satisfy the conditions $\dot{r}=0=\ddot{r}$, or equivalently, $V_{eff}(r_p)=0=\partial_rV_{eff}(r_p)$, such that $r_p$ is the radius of photon region. By the virtue of Eq. (\ref{veff1}), the circular light orbits must obey the conditions $\mathcal{R}(r_p)=0=\partial_r\mathcal{R}(r_p)$. The local maxima of the effective potential describes the unstable orbits that is mathematically expressed as $\partial^2_rV_{eff}(r_p)<0$.

The effective potential $V_{eff}$ has been plotted with respect to $r$ in Fig \ref{EP} showing the variation in the unstable orbits around the throat of the wormhole in comparison with Kerr and Kerr-Newman BHs. The upper left plot corresponds to the fixed values of $a$ and $b$, while $Q$ varies for each curve. For $Q=0$, the curve shows the light orbit outside the event horizon. Whereas, for other values of $Q$, the curves correspond to the orbits outside the wormhole throats. It is found that the unstable light orbit reduces in size with increase in $Q$ and each of which is attained at increasing effective potentials. Moreover, the deviation of unstable orbits and the respective effective potential for the wormhole in comparison with the Kerr BH is evident from the figure. The upper right plot corresponds to the fixed values of $a$ and $Q$, whereas $b$ varies for each curve. Contrary to the left plot, it can be deduced that the parameter $b$ is less sensitive because the variation in the curves is very small with increase in the value of $b$. However, it can be found that the size of unstable orbits increase with increase in the value of $b$ attained at decreasing values of effective potentials. The plot also shows the comparison of location of unstable orbits and effective potentials for the wormhole with the Kerr-Newman BH for $b=0$. The lower plot shows the variation of spin $a$ for each curve with fixed values of $b$ and $Q$. With increase in spin, the unstable orbits decrease in size, whereas each of such orbit is attained at increasing effective potential.

In order to calculate the shadow of the rotating wormhole in Bopp-Podolsky electrodynamics, we employ Bardeen's procedure for which the impact parameters can also be written in terms of $r_p$ as
\begin{figure}[t]
\centering
\subfigure{\includegraphics[width=0.35\textwidth]{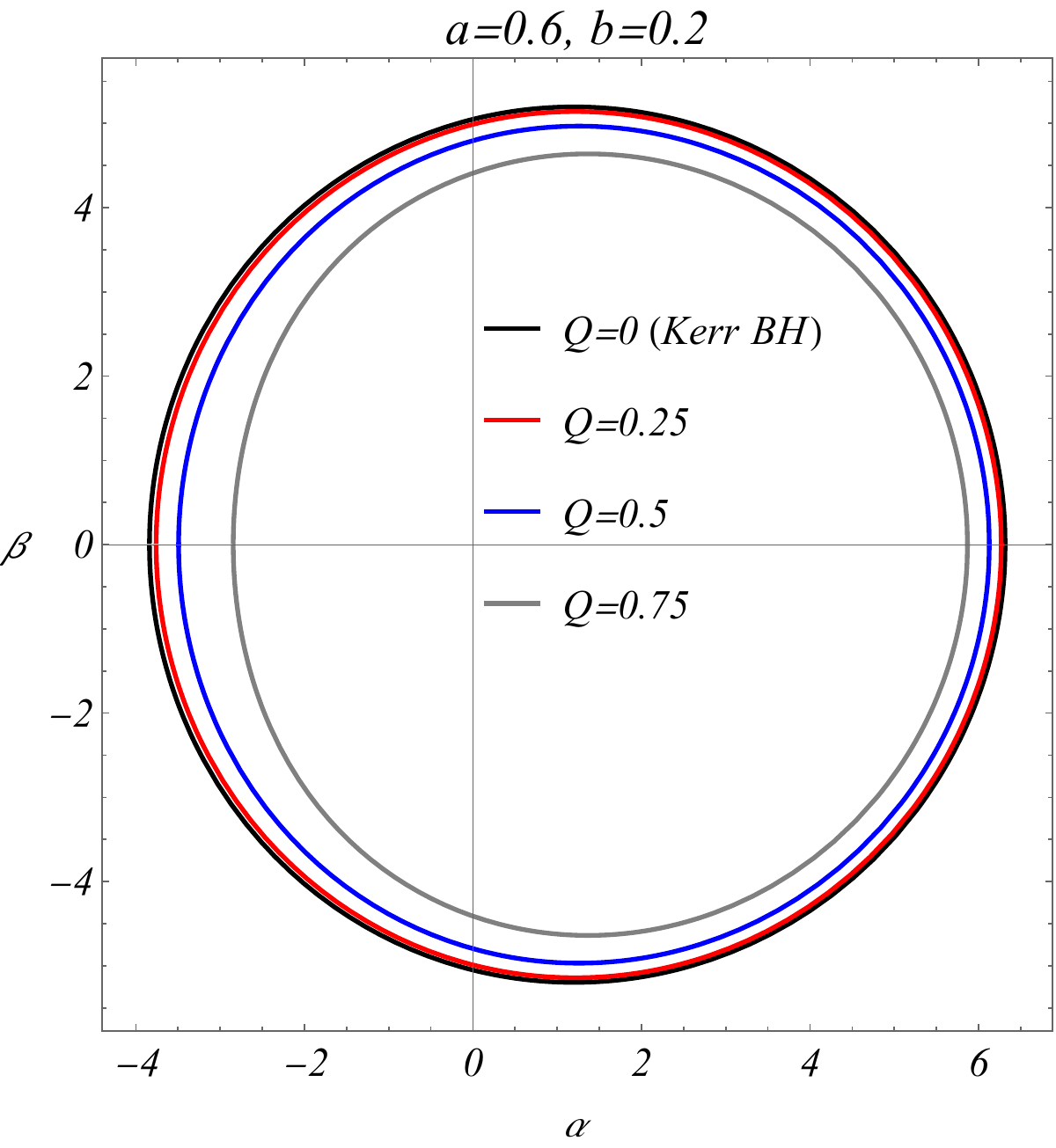}}~~~~~
\subfigure{\includegraphics[width=0.35\textwidth]{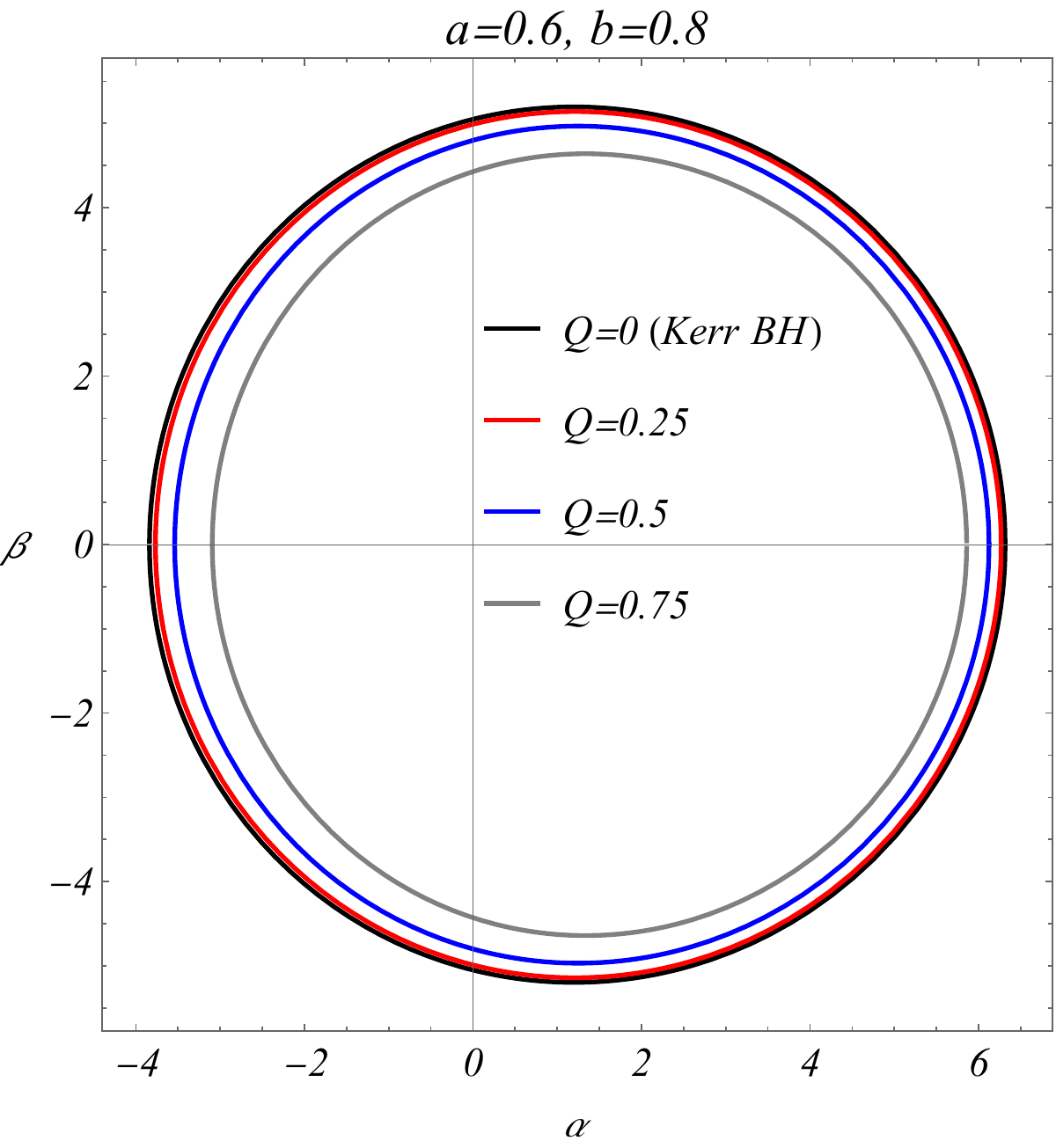}}\\
\subfigure{\includegraphics[width=0.35\textwidth]{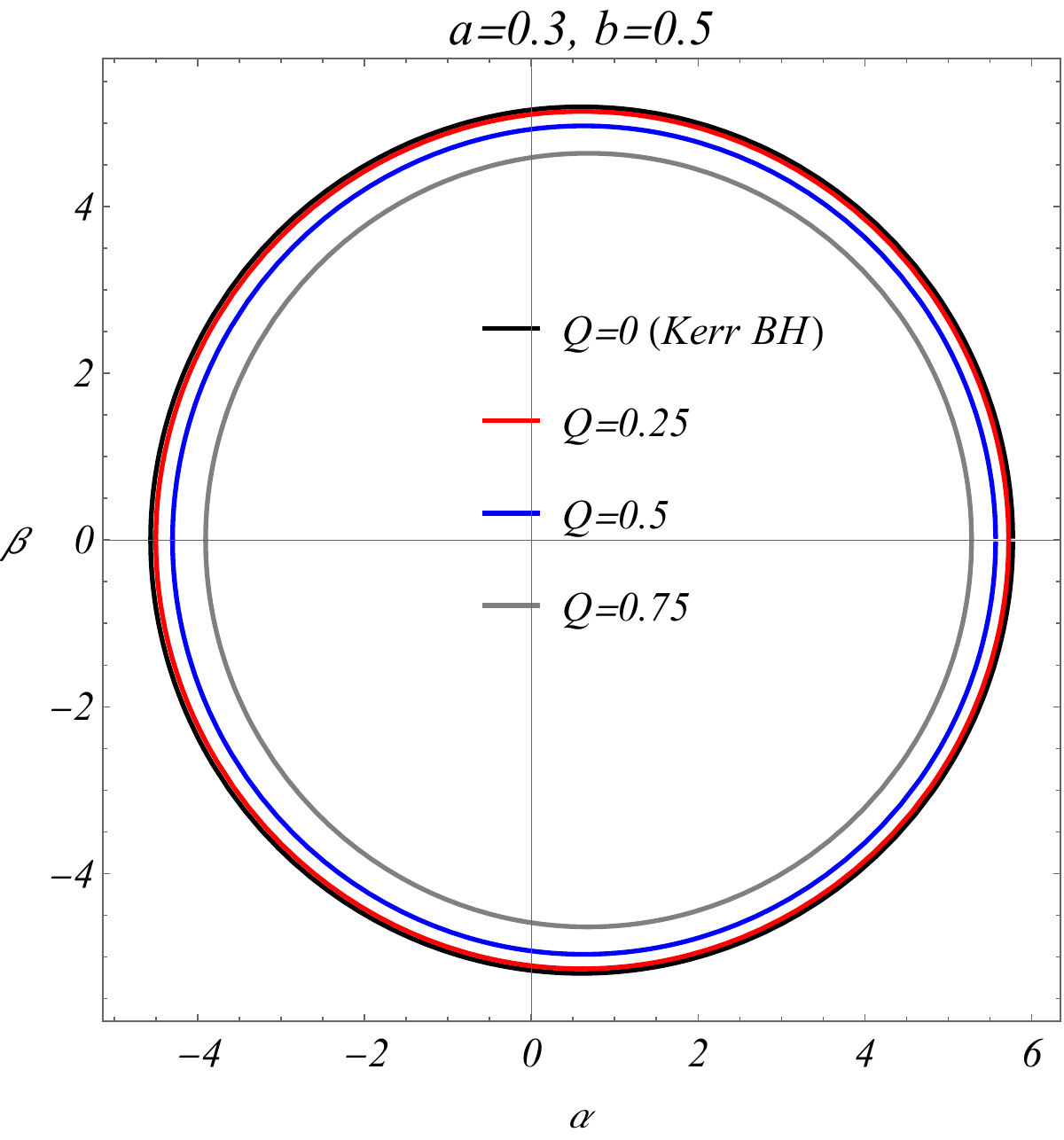}}~~~~~
\subfigure{\includegraphics[width=0.35\textwidth]{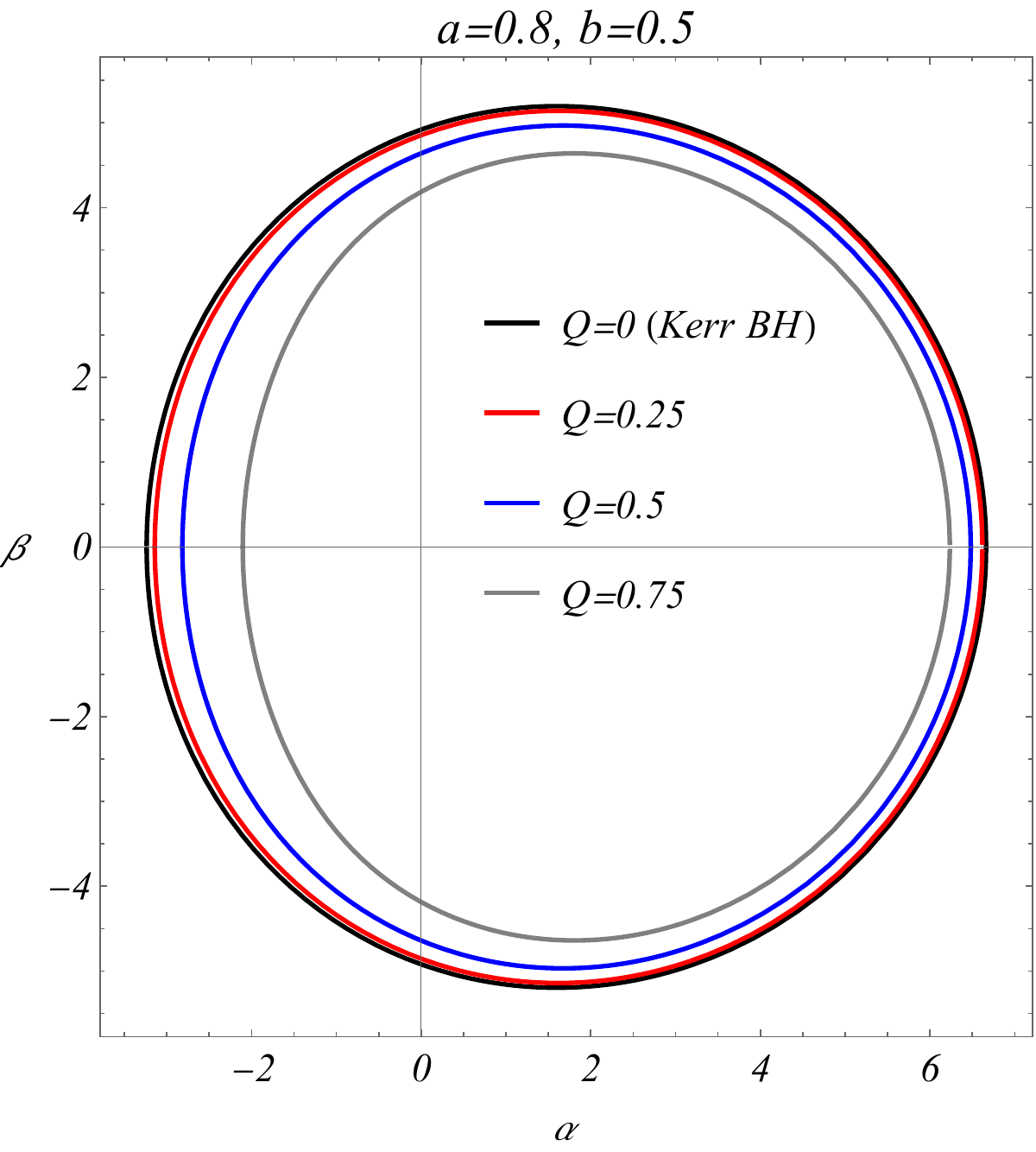}}
\caption{Variation of $Q$ for each curve showing the variation in shadows in celestial plane $\alpha$-$\beta$ for different values of $b$ and fixed $a$ in upper panel, and for different values of $a$ and fixed $b$ in lower panel with $M=1$. \label{Sha}}
\end{figure}
\begin{figure}[t]
\centering
\subfigure{\includegraphics[width=0.35\textwidth]{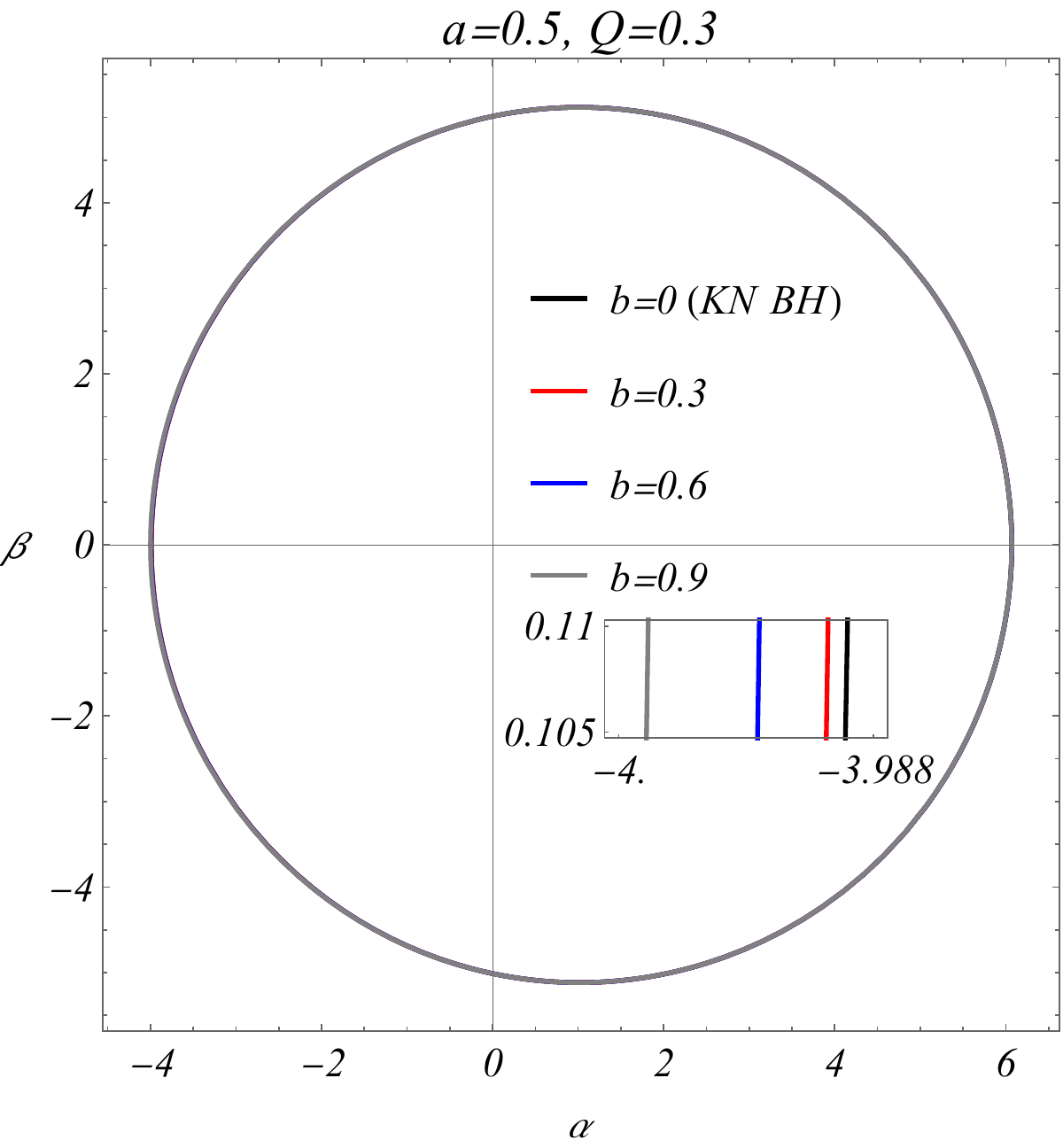}}~~~~~
\subfigure{\includegraphics[width=0.35\textwidth]{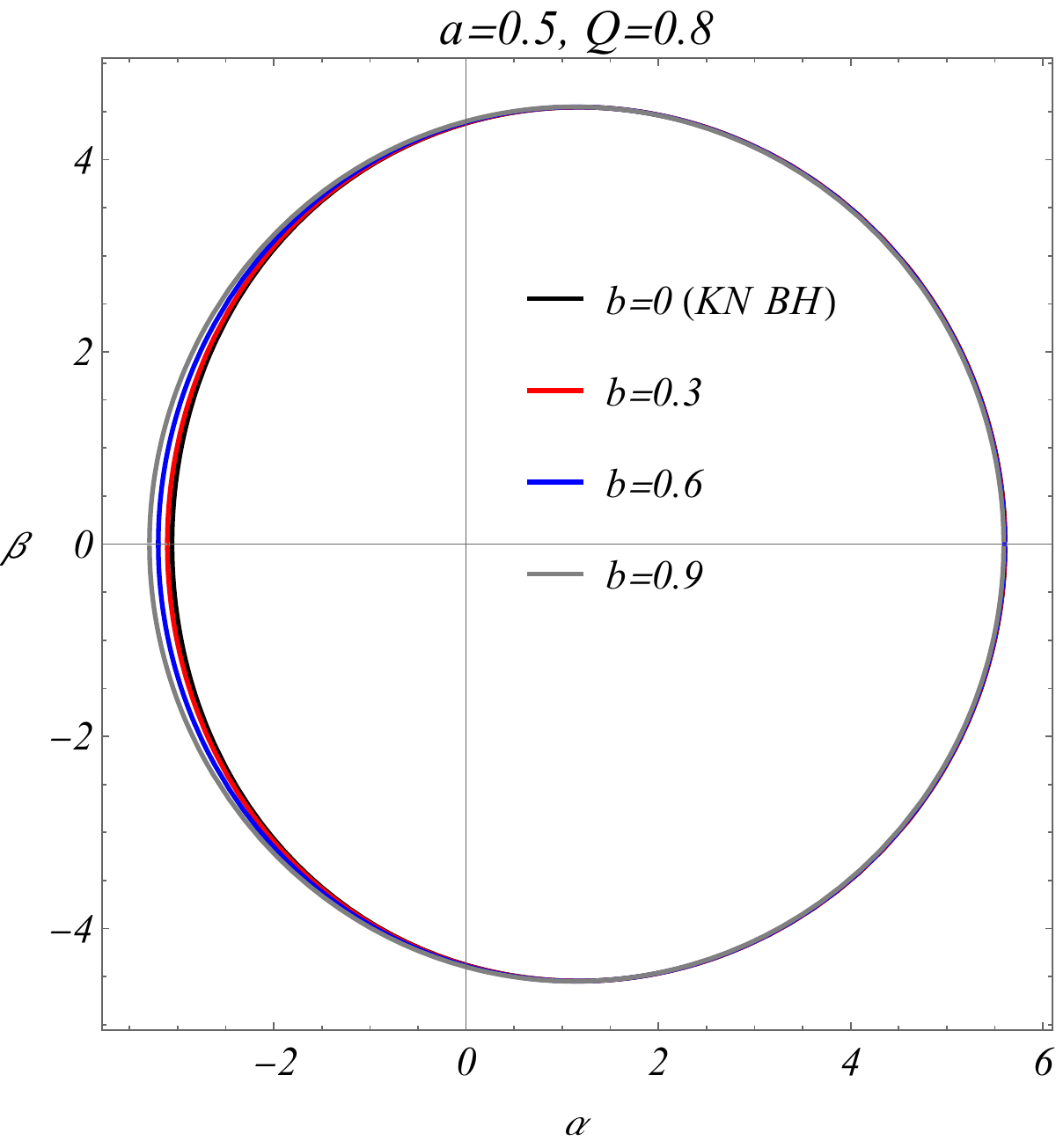}}\\
\subfigure{\includegraphics[width=0.35\textwidth]{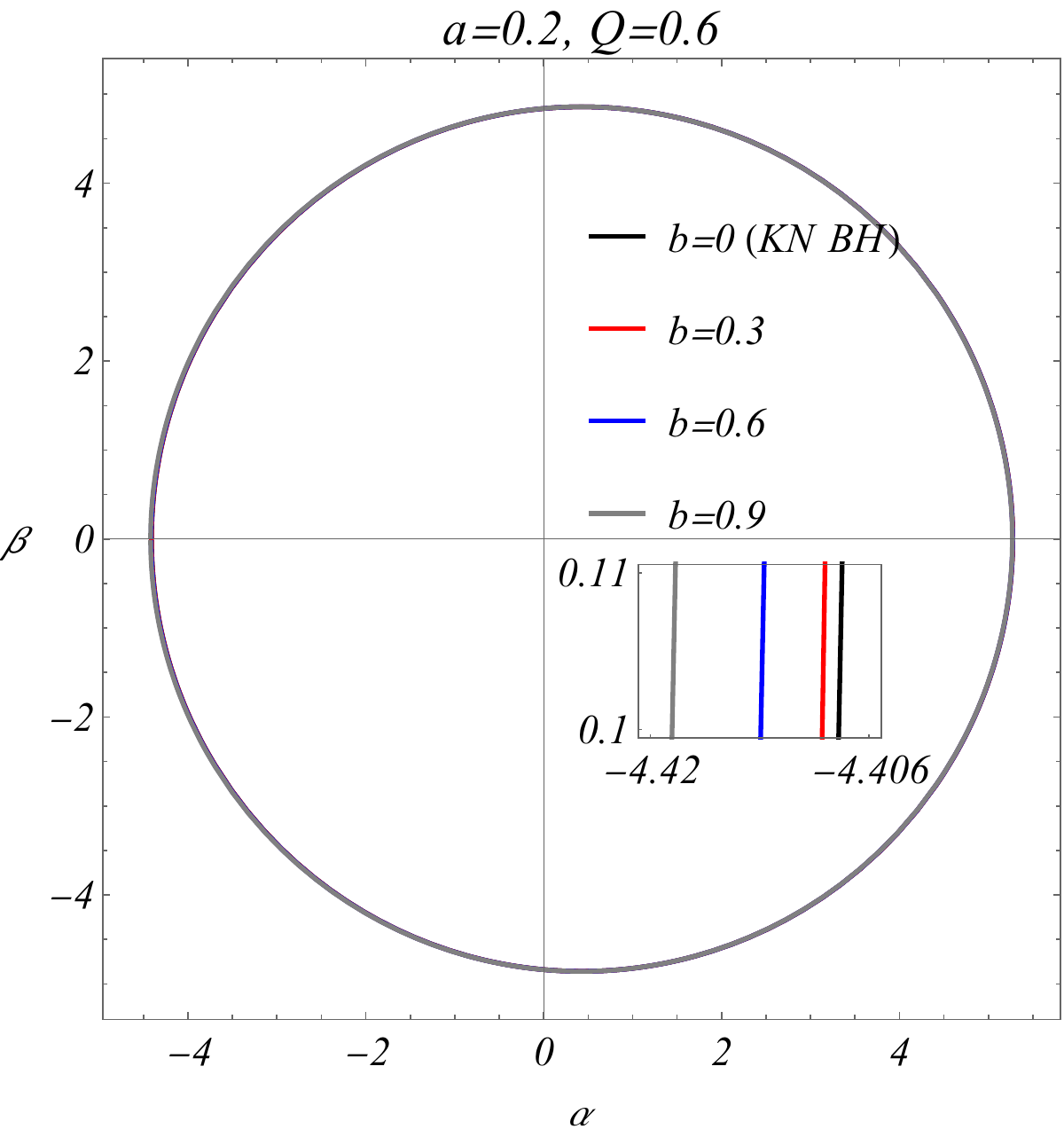}}~~~~~
\subfigure{\includegraphics[width=0.35\textwidth]{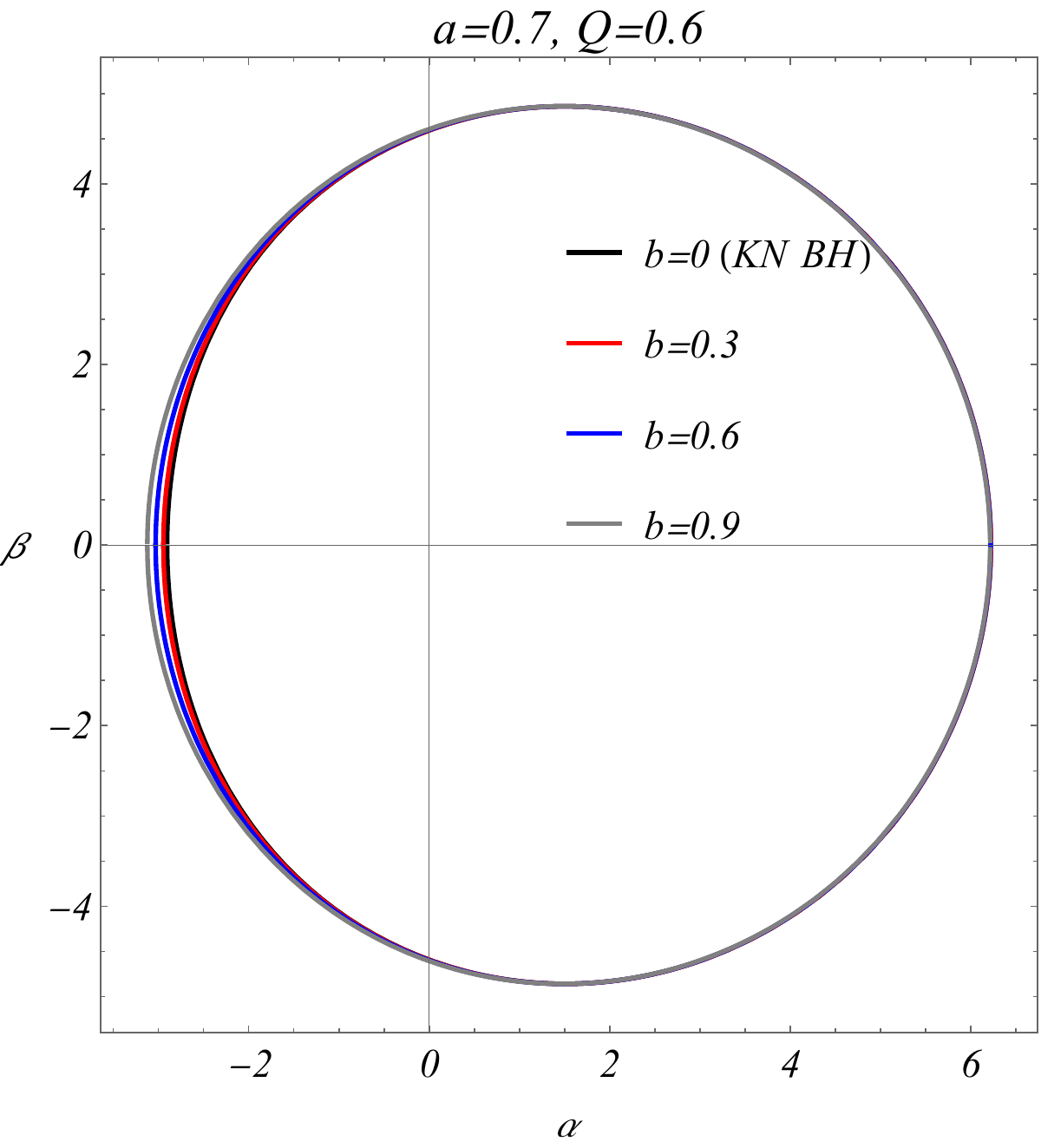}}
\caption{Shadow plots for different values of $b$ corresponding to each curve while keeping $a$ fixed and varying $Q$ in upper panel, and keeping $Q$ fixed and varying $a$ in lower panel with $M=1$. \label{Shb}}
\end{figure}
\begin{figure}[t]
\centering
\subfigure{\includegraphics[width=0.35\textwidth]{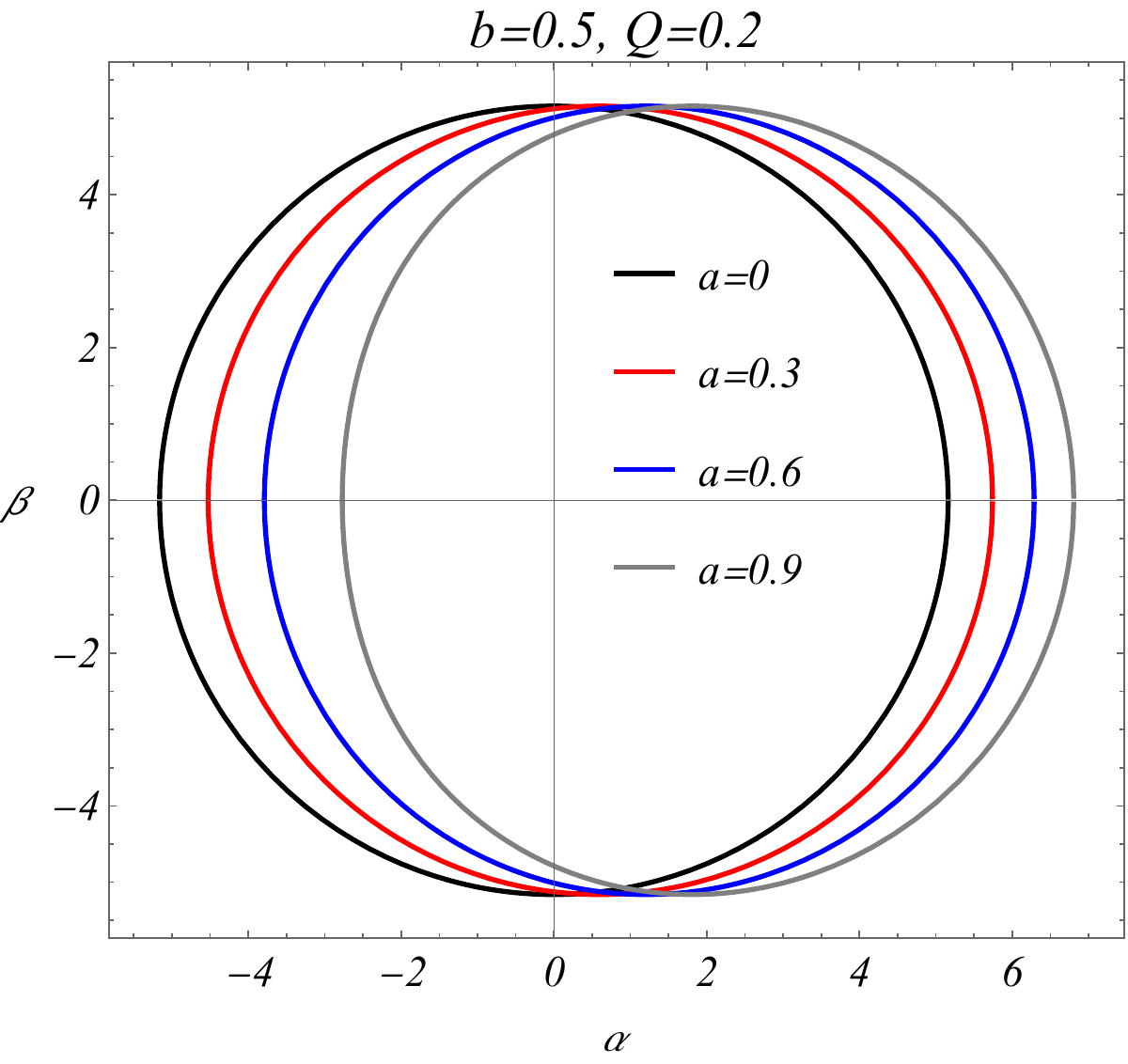}}~~~~~
\subfigure{\includegraphics[width=0.35\textwidth]{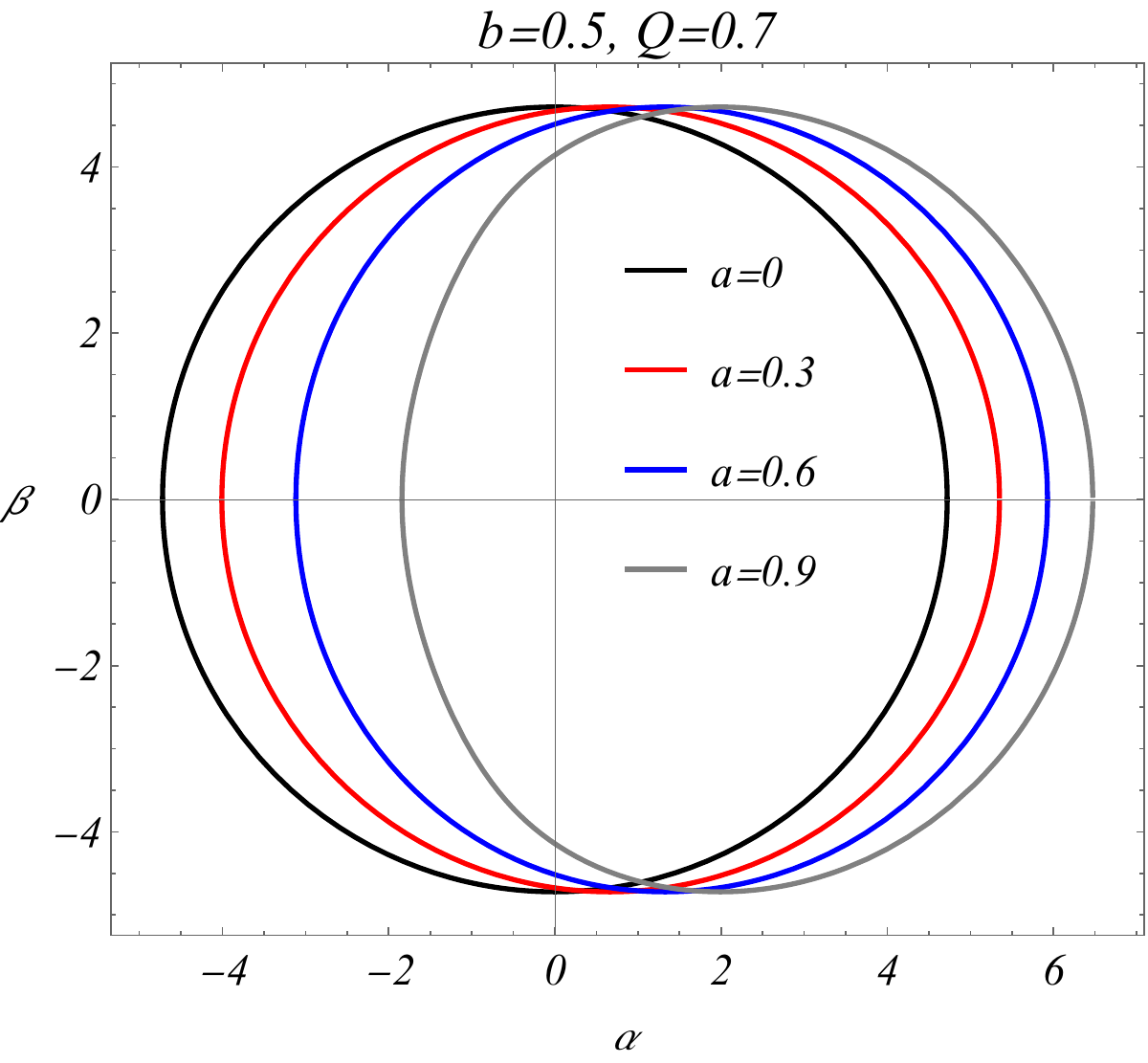}}\\
\subfigure{\includegraphics[width=0.35\textwidth]{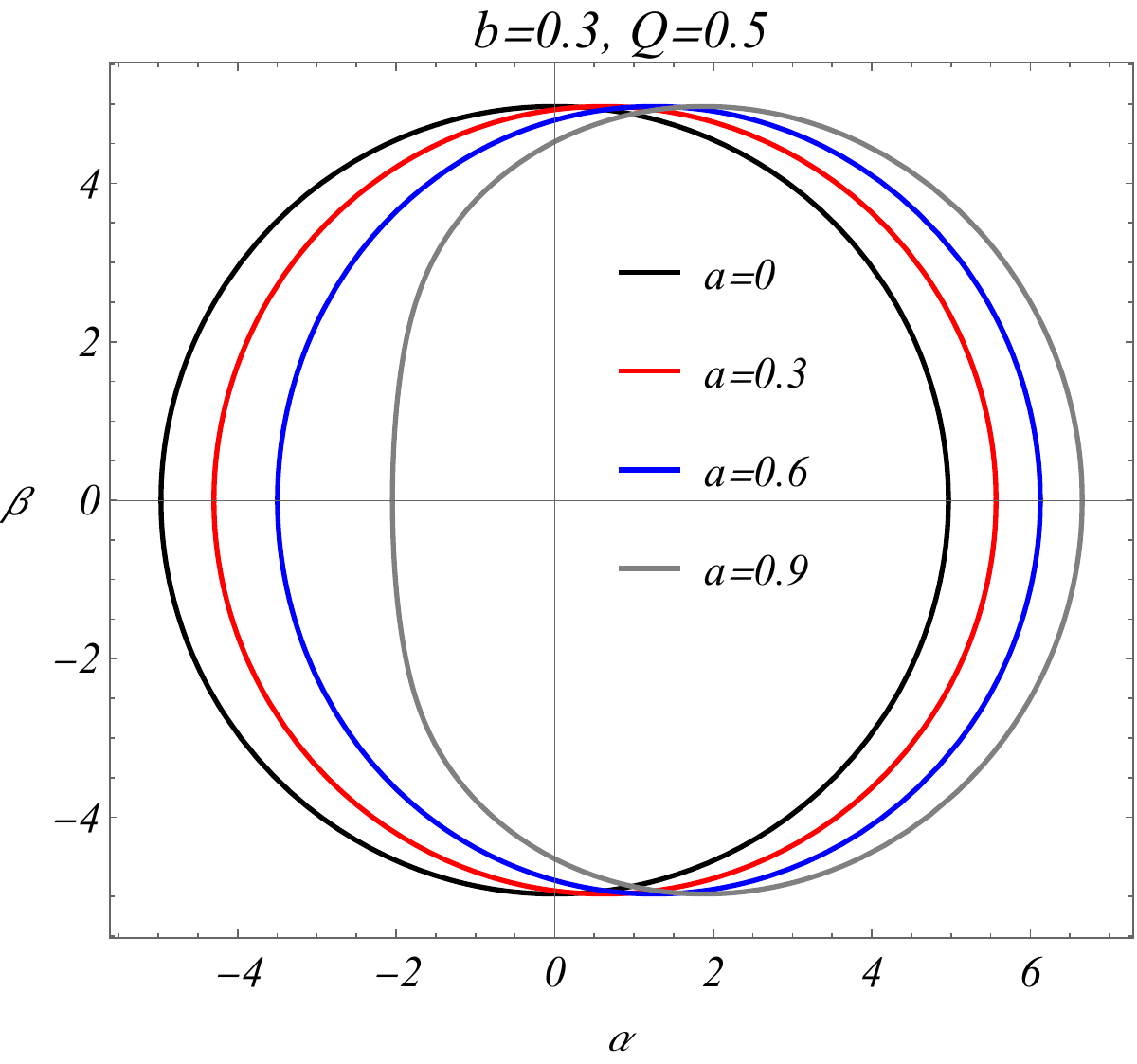}}~~~~~
\subfigure{\includegraphics[width=0.35\textwidth]{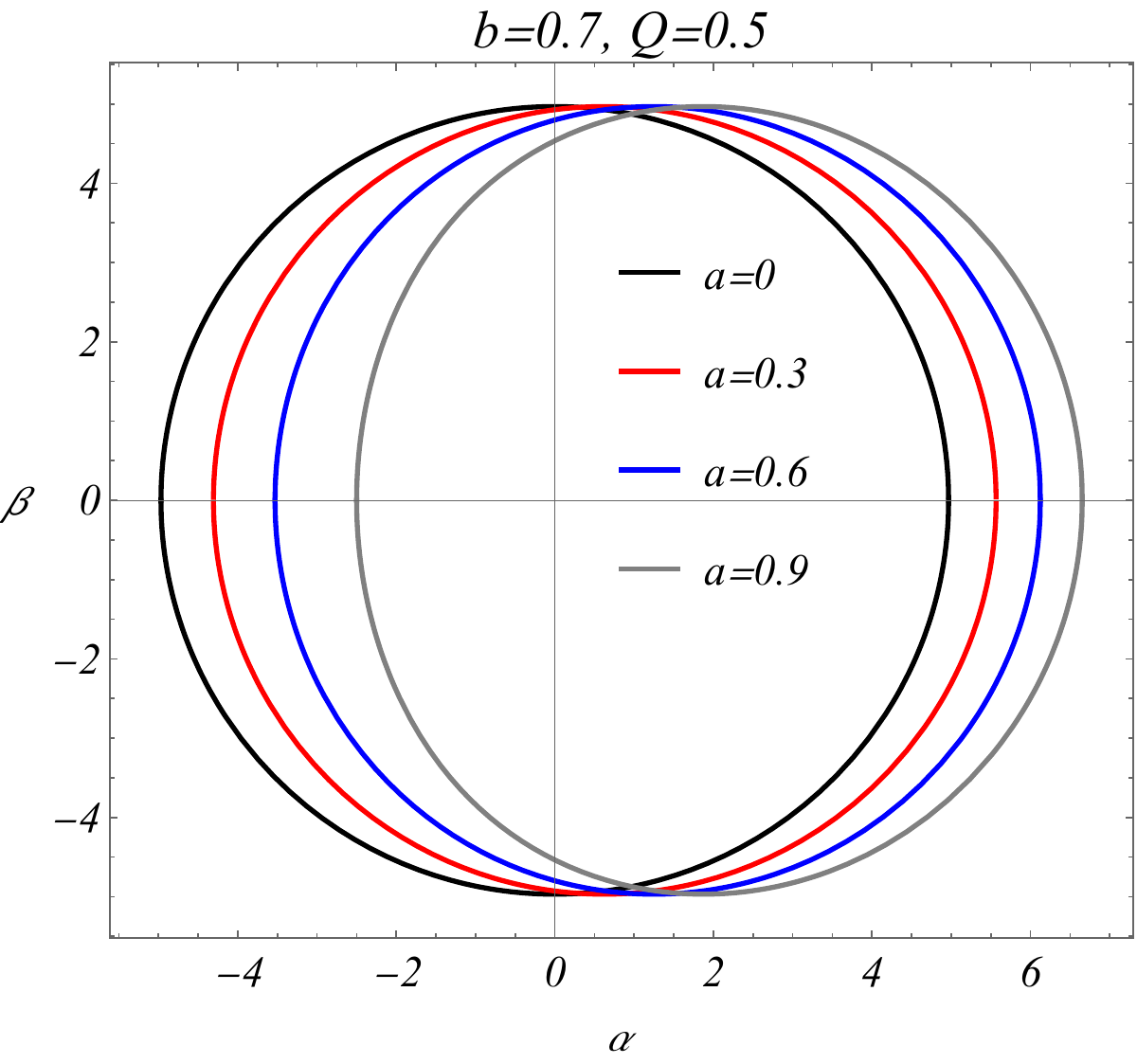}}
\caption{Behavior of $a$ corresponding to each shadow curve for different $Q$ and fixed $b$ in upper panel, and different $b$ and fixed $Q$ with $M=1$. \label{Shc}}
\end{figure}
\begin{eqnarray}
u(r_p)&=&\frac{r_p^2+a^2}{a}-\frac{4r_p\Delta(r_p)}{a\Delta'(r_p)}, \label{ipa} \\
\Sigma(r_p)&=&\frac{r_p^2\left(8r_p\Delta(r_p)\Delta'(r_p)-16\Delta(r_p)\left(\Delta(r_p)-a^2\right)-r_p^2\left(\Delta'(r_p)\right)^2\right)}{a^2\left(\Delta'(r_p)\right)^2}. \label{ipb}
\end{eqnarray}
The new definitions for the impact parameters may also be considered as $L_E=\frac{L}{E}$ and $K_E=\frac{\mathcal{Z}+(l-aE)^2}{E^2}$ that are explicitly written in terms of $r_p$ as
\begin{eqnarray}
K_E(r_p)&=&\frac{16r_p^2\Delta(r_p)}{(\Delta'(r_p))^2}, \label{KE}\\
L_E(r_p)&=&u(r_p)=\frac{a^2+r_p^2}{a}-\frac{4r_p\Delta(r_p)}{a\Delta'(r_p)} \label{LE}
\end{eqnarray}
that are essentially helpful in calculating the shadows for static cases. For the rotating wormhole, we may calculate the shadows by incorporating the parametric equations with $r_p\in\mathcal{I}=\left[r_{p,min},r_{p,max}\right]$ as the parameter having upper and lower bounds. These boundary points of $r_p$ can be calculated by solving the equation $\Sigma(r_p)=0$ for the positive real roots. If the wormhole is non-rotating, the value of $r_p$ and hence $K_E(r_p)$ is unique so that $r_p$ cannot be treated as a parameter. Moreover, setting $a=0$ for the static case, we cannot calculate the value of $L_E(r_p)$ as it becomes indeterminate. Therefore, it remain unknown and hence we may treat it as a parameter. Note that its allowed values are positive real numbers in an interval with upper and lower bounds as the roots determined by the equation $\Theta(\pi/2)=0$. The 3D wormhole is projected onto the celestial plane $\alpha$-$\beta$ as a shadow image by the relations
\begin{eqnarray}
\alpha(r_p)&=&-\lim\limits_{r\rightarrow\infty}\bigg[\frac{d\phi}{dr}r^2\sin\theta\bigg]_{\theta\rightarrow\theta_0}, \label{alpha}\\
\beta(r_p)&=&\lim\limits_{r\rightarrow\infty}\bigg[\frac{d\theta}{dr}r^2\bigg]_{\theta\rightarrow\theta_0}, \label{beta}
\end{eqnarray}
where, the observer's location according to the symmetry of the spacetime is given by the angle $\theta_0$. Using the geodesic equations, the celestial parameters in the Eqs. (\ref{alpha}) and (\ref{beta}) become
\begin{eqnarray}
\alpha(r_p)&=&-u(r_p)\csc\theta_0, \label{alp1}\\
\beta(r_p)&=&\pm\sqrt{\Sigma(r_p)+a^2\cos^2\theta_0-u(r_p)^2\cot^2\theta_0} \label{bet1}
\end{eqnarray}
and since, the shadow is observed at the inclination $\left(\theta=\frac{\pi}{2}\right)$, therefore, the celestial parameters are reduced to
\begin{eqnarray}
\alpha(r_p)&=&-u(r_p), \label{alp2}\\
\beta(r_p)&=&\pm\sqrt{\Sigma(r_p)}. \label{bet2}
\end{eqnarray}

We have considered different cases for the parameters $a$, $b$ and $Q$ for obtaining the optical shadow images of the rotating wormhole in Bopp-Podolsky electrodynamics that have been plotted in the Figs. \ref{Sha}, \ref{Shb} and \ref{Shc}. In Fig. \ref{Sha}, the variation in the shadow size and shape has been depicted with respect to the variation in $Q$ corresponding to each curve. It can be found that for each plot, the shadow shrinks as $Q$ increases. The case $Q=0$ corresponds to the Kerr BH which has the largest shadow size among all other curves. From which we can infer that the inclusion of the charge $Q$ reduces the size of the shadow image of the wormhole and transforms it to the wormhole for $Q\neq0$ as being a BH for the case $Q=0$. Moreover, increasing the values of $b$ in the upper panel and $a$ in the lower panel, mainly affect the distortion in the shadows that is obvious from the Fig. \ref{Sha}. In Fig. \ref{Shb}, the variation of $b$ corresponding to each curve shows a rigid behavior in the variation of the shadow size, that is, the shadow size is not much altered by varying the parameter $b$. However, the shapes of the shadows vary for some cases as can be seen in the right panel. The case $b=0$ corresponds to the Kerr-Newman BH for which the distortion in the shadows is the highest among all other curves that correspond to the wormhole case for $b\neq0$. Therefore, the parameter $b$ reduces the flatness in the shadow images of the wormhole which is more visible for the large values of $Q$ and $a$ in the right panel. In Fig. \ref{Shc}, a comparison of the static and the rotating wormhole shadows has been generated for the fixed values of $b$ and $Q$ in each plot. It can be seen that the static wormhole shadow is an exact circle, whereas, with the increase in spin $a$, the shadows are shifted towards the right and with increased flatness. Since this property of spin parameter in BHs is due to the existence of an ergosphere. However, in this study, despite the rotating wormhole (\ref{31}) does not have any ergosphere, still the shadows are flattened due to increase in the spin parameter. Unlike BHs, the rotating wormhole (\ref{31}) does not have an extremal spin value. However, there exist an upper limit $a=a^{crit}$ that can be termed as critical value of spin, above which the shadows are not formed because the curves obtained are not closed.

Now, we discuss the shadows of rotating wormhole (\ref{31}) for critical value of spin. As said before, there exist an extremal value of spin in case of BHs above which there exist a naked singularity instead of an event horizon. However, for the wormhole (\ref{31}), we can determine a critical value of spin above which shadows are not formed as closed curves. We consider it as maximum allowed value of spin in our case. For rotating wormholes, the unstable photon orbits are divided into two distinct classes for critical and near critical values of spin. That is, the inner unstable orbits located at the wormhole throat, and the outer unstable orbits that are located radially outside the throat. The wormhole solution (\ref{31}) generates cuspy shadows for certain values of spin. Since, the merging of the two families of unstable orbits is not smooth for critical and near critical spin values. Therefore, these cusps are formed at the intersection of the shadow curves determined from the above mentioned classes of unstable null orbits. In fact, for lower values of spin, the shadows are formed by the contribution of only outer unstable orbits for which there does not exist cusps and the shadows are smooth curves as can be seen in Fig. \ref{Shc}. The family of unstable null orbits lying outside the throat of the wormhole is obtained from the local maxima of the effective potential such that the function $g^{rr}\neq0$ and the conditions
\begin{eqnarray}
\mathcal{R}(r)=0, \quad \partial_r\mathcal{R}(r)=0, \quad \partial^2_r\mathcal{R}(r)\geq0 \label{cond1}
\end{eqnarray}
must be satisfied. The other family of unstable null orbits are located at the wormhole throat that are determined from the local maxima of the effective potential such that the function $g^{rr}=0$ and the conditions
\begin{eqnarray}
1-\frac{b(r)}{r}=0, \quad \mathcal{R}(r)=0, \quad \partial_r\mathcal{R}(r)\geq0 \label{cond2}
\end{eqnarray}
are satisfied. The left branch of the shadow curve is located at the throat $r_0$, whereas, the right branch is located well outside the throat at some radial distance $r>r_0$. Due to the presence of disjoint families of orbits, the cusps are formed in the shadows. The critical value of spin can be determined by solving the system
\begin{figure}[t]
\centering
\subfigure{\includegraphics[width=0.28\textwidth]{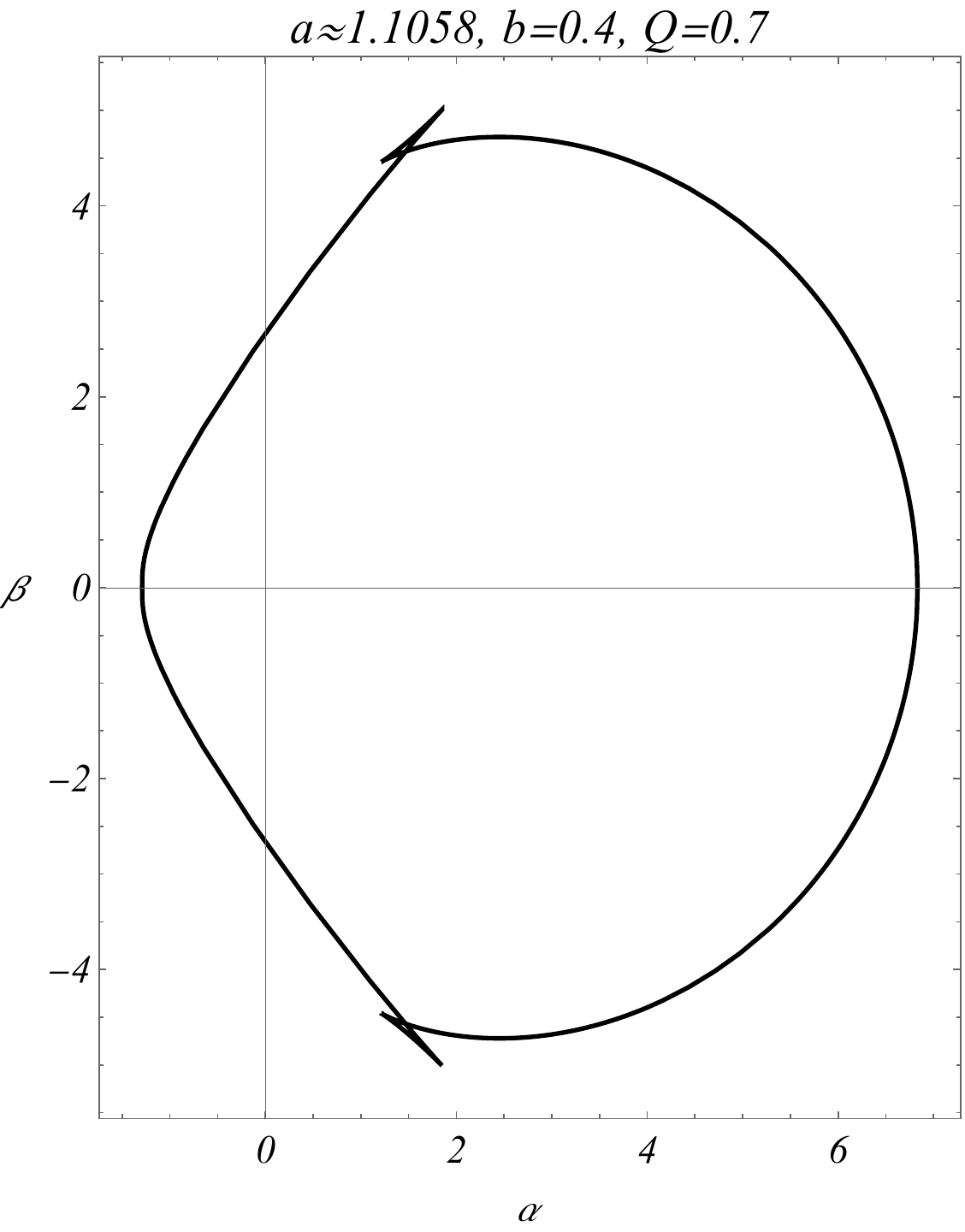}}~~
\subfigure{\includegraphics[width=0.28\textwidth]{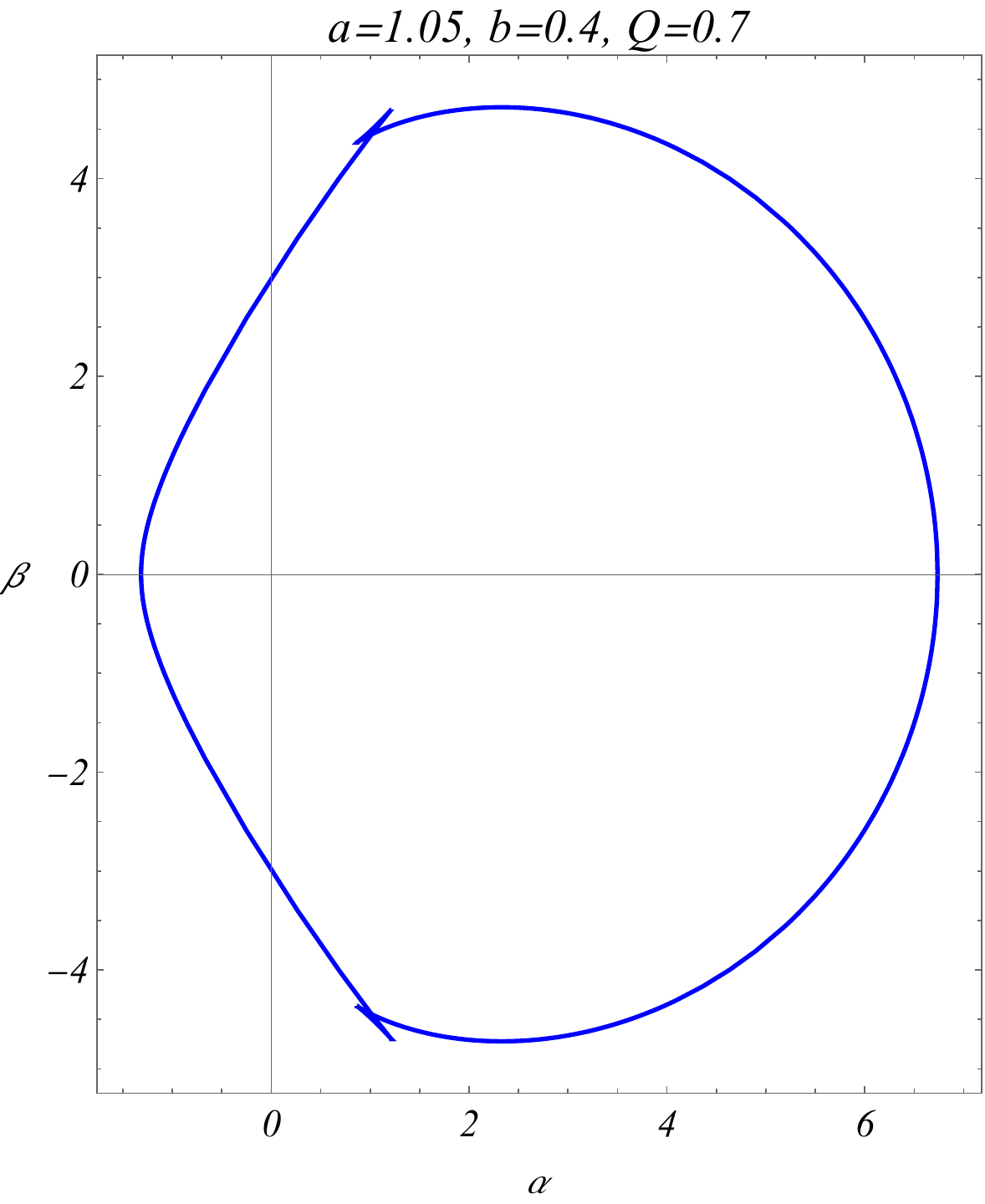}}~~
\subfigure{\includegraphics[width=0.28\textwidth]{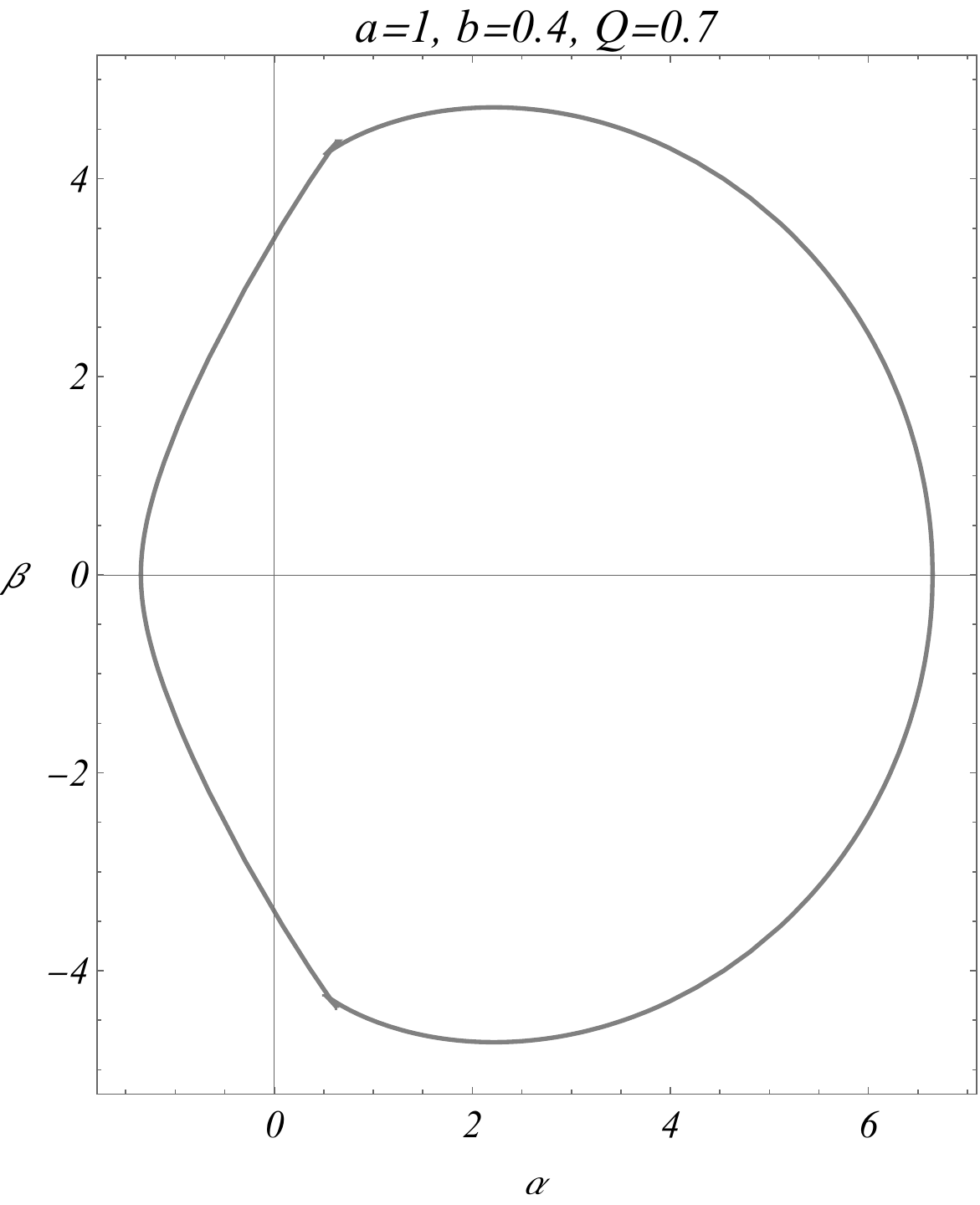}}
\caption{The plots showing the shadows with cusps for critical value of spin and near critical values for fixed values of $b$ and $Q$ while keeping $M=1$. \label{Csp1}}
\end{figure}
\begin{figure}[t]
\centering
\subfigure{\includegraphics[width=0.4\textwidth]{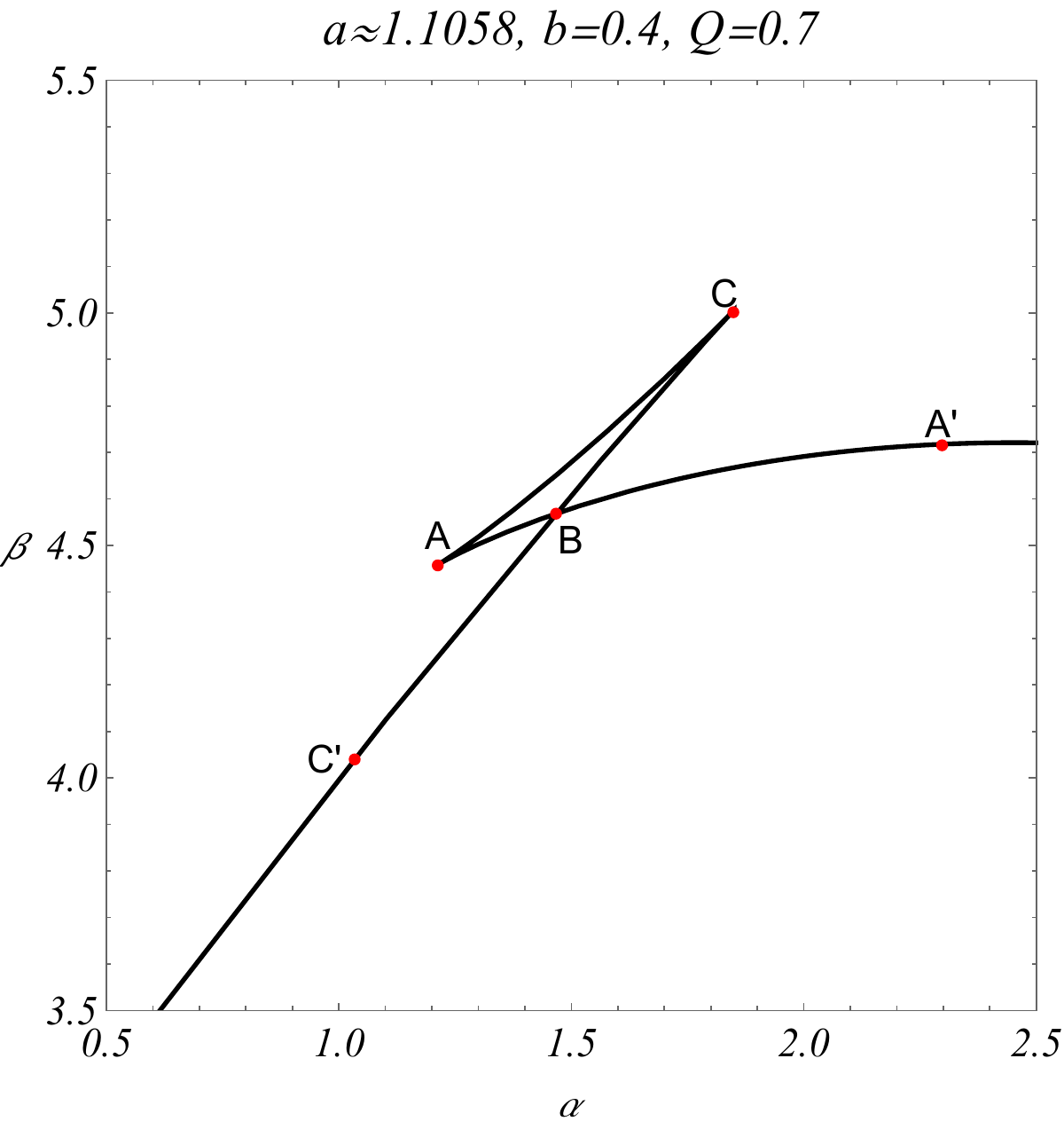}}
\caption{A detailed plot for cuspy part of the shadow in the left plot in Fig. \ref{Csp1} highlighting the intersection point and the stability points. \label{Csp2}}
\end{figure}
\begin{eqnarray}
\beta(r)=0, \quad \partial_r\beta(r)=0 \label{crit}
\end{eqnarray}
for some fixed values of $b$ and $Q$. We have calculated it for $b=0.4$ and $Q=0.7$ that comes out to be $a=a^{crit}\approx1.1058$. The shadows obtained for this particular case and for some near critical values of spin have been plotted in Fig. \ref{Csp1} that clearly show the formation of cusps at the shadow boundaries. The interface of both branches of shadows is more clearly visualized in zoomed-in plot in Fig. \ref{Csp2} corresponding to the left-most plot in the Fig. \ref{Csp1}. We found that the branch of the shadow curve generated by the unstable orbits residing at the throat is defined for both positive and negative values of the celestial coordinate $\alpha$. However, the branch of the shadow curve determined by the unstable photon orbits outside the throat is defined only for the positive values of $\alpha$. Therefore, the junction of both branches of the curves is located at some $\alpha>0$. From the Fig. \ref{Csp1}, it can be inferred that for $a=a^{crit}\approx1.1058$, the cusps are formed at the largest possible value of $\alpha$ and when the value of $a$ decreases, the cusps are shifted towards the axis $\alpha=0$. Moreover, as the value of spin decreases from the maximum value $a=a^{crit}\approx1.1058$, the cusps are also reduced in size and thus are diminished for some value of $a<a^{crit}$ generating smooth shadows that are ultimately due to the outer unstable orbits only. A detailed image of the cusps are shown in Fig. \ref{Csp2}. The branch of the shadow curve associated with the outer family of unstable orbits is denoted by $A'A$ and turns stable at point $A$ where a cusp is formed. The other branch associated with the inner family of photon orbits at the throat is denoted by $C'C$ having the stability point denoted by $C$ where another cusp is formed. Both branches intersect each other at point $B$.

\section{Strong Gravitational Lensing}\label{sec5}
\begin{figure}[t]
\centering
\subfigure{\includegraphics[width=0.45\textwidth]{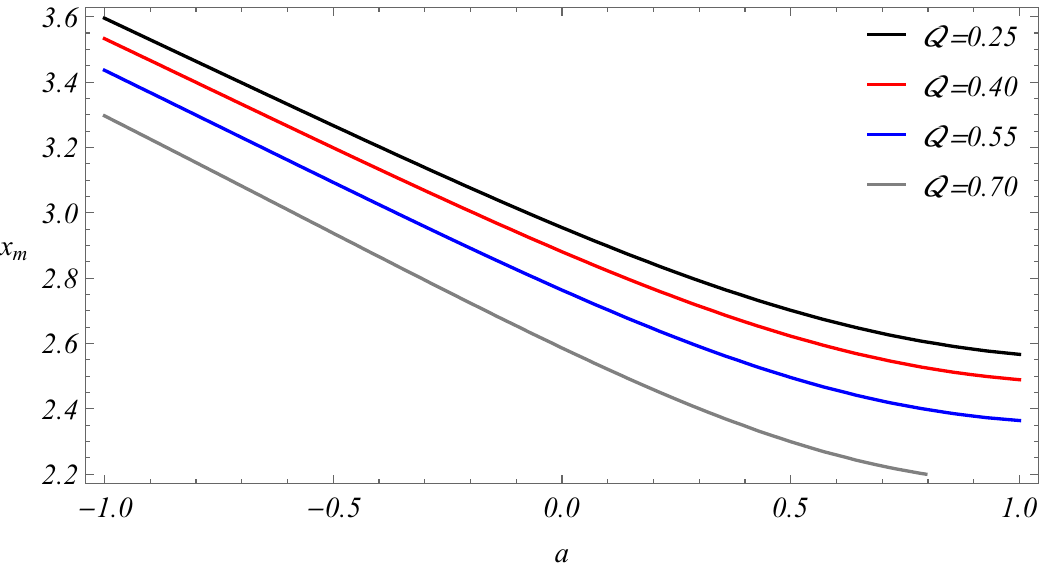}}~~~
\subfigure{\includegraphics[width=0.45\textwidth]{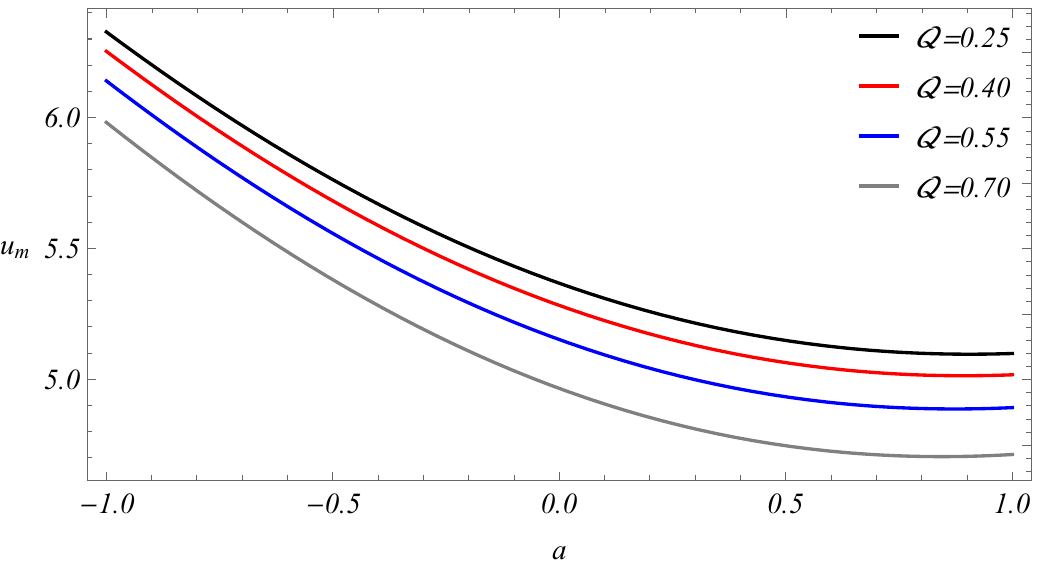}}
\subfigure{\includegraphics[width=0.45\textwidth]{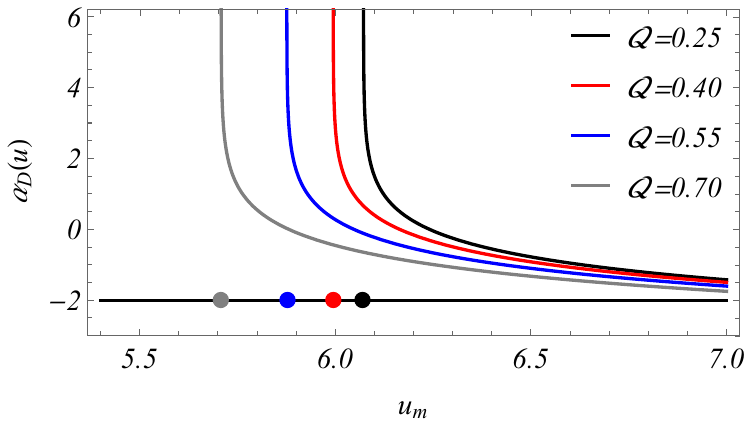}}~~~
\subfigure{\includegraphics[width=0.45\textwidth]{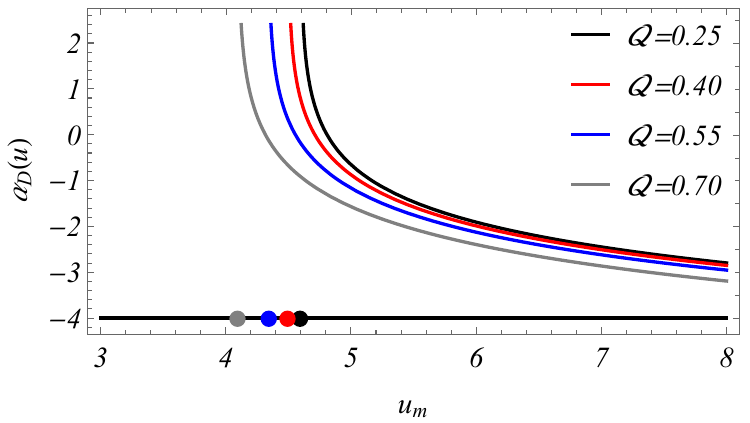}}
\subfigure{\includegraphics[width=0.45\textwidth]{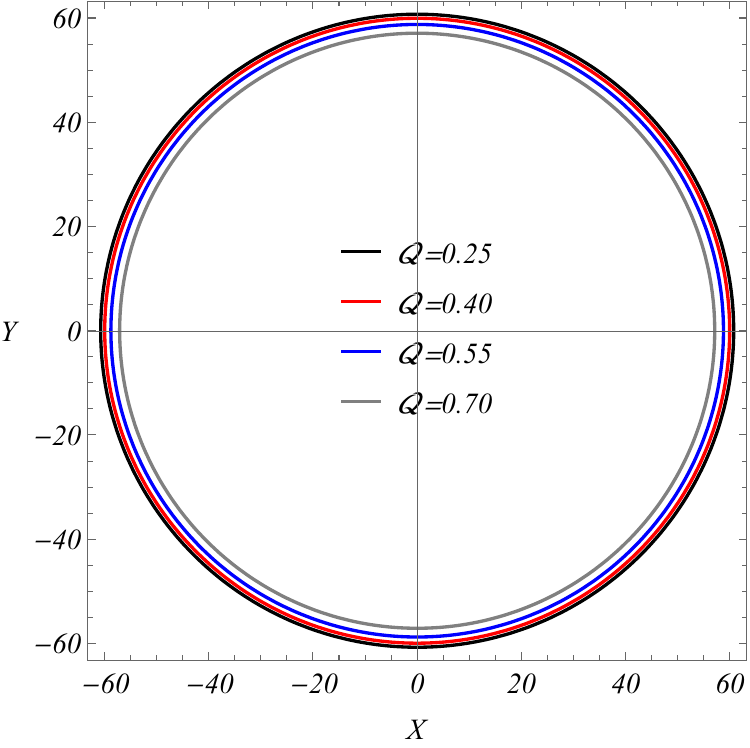}}~~~
\subfigure{\includegraphics[width=0.45\textwidth]{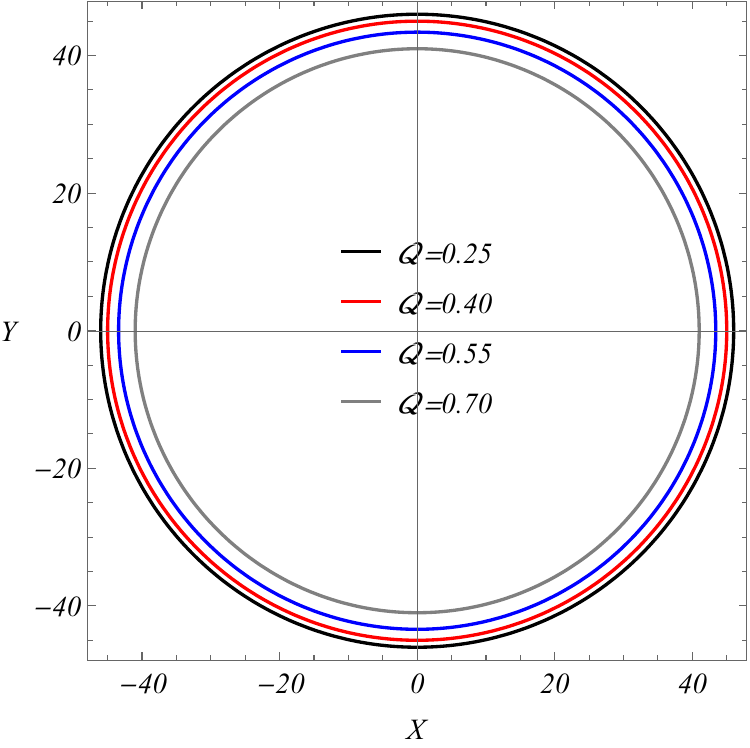}}
\caption{\textbf{Top row}: The left panel shows the trend of the radius photon sphere $x_m$ versus the rotation parameter $a$, while the right panel is depicting the the impact parameter $u_m$ against the rotation parameter $a$. \textbf{Middle row}: The left and right panels are illustrating the trend of the deflection angle $\alpha_D(u_m)$ in the strong field regime versus the impact parameter $u_m$. In this case, the rotation parameter has been fixed to be $a=-0.8$ (left panel) and $a=0.8$ (right panel). \textbf{Bottom row}: The left and right panels are exhibiting the angular position of the outermost Einstein’s ring ($\theta^{\text{E}}_1$). In this case, the rotation parameter has been fixed to be $a=-0.8$ (left panel) and $a=0.8$ (right panel). For all these plots we have considered $M=1$ and $b=0.5$. \label{fig8}}
\end{figure}

In this section, we study the strong gravitational lensing produced by the rotating wormhole (\ref{27}). However, the lensing phenomenon for wormholes is different from the BHs because the photon orbit radius or photon sphere is not uniquely defined. Namely, as was studies in \cite{h_2019aaa} for the static case (it is not the exception for the rotating one), the strong gravitational lensing for wormholes takes into account the following situations:
\begin{itemize}
\item \textbf{a.} The observer and the source are on the same side of the wormhole throat. Three sub-cases may also arise here:\\
\textbf{a1.} Strong lensing occurs due to the presence of a photon sphere outside the throat.\\
\textbf{a2.} The throat itself acts as a photon sphere in strong lensing.\\
\textbf{a3.} Strong lensing occurs due to both a photon sphere and an anti-photon sphere.
\end{itemize}
\begin{itemize}
\item \textbf{b.} The observer and the source are on opposite sides of the throat, that is, the observer sees the light coming from another universe. Two distinct sub-cases may take place:\\
\textbf{b1.} A photon sphere outside the throat is involved in strong lensing.\\
\textbf{b2.} The throat itself acts as a photon sphere in strong lensing.
\end{itemize}
As can be seen, the cases \textbf{a1} and \textbf{a2} are the same as the cases \textbf{b1} and \textbf{b2} (see \cite{h_2019aaa} for further details). Keeping in view the above situations, we intend to consider the case when the observer and the source are on opposite sides of the throat, besides the photon sphere is outside the throat, that is, $r_0<r_p$. For this, we follow the analysis presented in \cite{Bozza:2002zjaaa,Bozza:2008evaaa}. Therefore, to begin with, we restrict to the equatorial plane at $\theta=\frac{\pi}{2}$ which implies $\dot{\theta}=0$. This allows us to express the most general rotating line element as 
\begin{equation}
ds^2=-\bar{A}(x)dt^2+\bar{B}(x)dx^2+\bar{C}(x)d\phi^2-\bar{D}(x)dtd\phi, \label{gmet}
\end{equation}
where, for this particular case, the coefficients $\{\bar{A},\bar{B},\bar{C},\bar{D}\}$ can be considered as in Eq. (\ref{31}). For rotating bodies, the deflection angle is given by \cite{Bozza:2002afaaa} 
\begin{equation}\label{angle}
\alpha_D\left(x_0\right)=2\int_{x_0}^{\infty}\frac{d\phi}{dx}dx=2\int_{x_0}^{\infty} \frac{\sqrt{\bar{A}_0\bar{B}}\left(2\bar{A}L+\bar{D}\right)}{\sqrt{4\bar{A}\bar{C}+\bar{D}^2} \sqrt{\bar{A}_0\bar{C}-\bar{A}\bar{C}_0+L\left(\bar{A}\bar{D}_0-\bar{A}_0\bar{D}\right)}}dx-\pi,
\end{equation}
with $\{\bar{A}_0,\bar{C}_0,\bar{D}_0\}=\{\bar{A}(x_0),\bar{C}(x_0),\bar{D}(x_0)\}$, that is, the coefficients are evaluated at the distance of closest approach. Besides, $L$ represent the projection of the angular momentum. In the case of strong deflection limit (SDL), the argument of the integral (\ref{angle}) is expanded near the photon sphere\footnote{As we are following the procedure given in \cite{Bozza:2002zjaaa,Bozza:2002afaaa}, from now on the photon sphere radius $r_p$ will be denoted by $x_m$.} $r_p=x_m$. As it is well-known (see also the discussion provided in Section \ref{section4}), from the radial geodesic equation the following conditions hold $V_{eff}(x_m)=0=\partial_rV_{eff}(x_m)$ at the photon sphere locus. These conditions allow us to get two important equations, one defining the photon sphere radius given by
\begin{equation}\label{photonra}
A(x)C'(x)-A'(x)C(x)+u_0\left[A'(x)D(x)-A(x)D'(x)\right]=0
\end{equation}
and the second one is the so-called impact parameter $u\equiv\frac{L}{E}$ 
\begin{equation}\label{impact}
u_0=L=\frac{-\bar{D}_0+\sqrt{4\bar{A}_0\bar{C}_0+\bar{D}_0^2}}{2\bar{A}_0},
\end{equation}
where, the energy has been fixed to be $E=1$. It is worth mentioning that the photon sphere corresponds to the critical impact parameter. After replacing the expression (\ref{impact}) into the equation (\ref{photonra}), one gets an equation for determining the radius of the photon sphere independent of the impact parameter $u_0$. As we are interested in the case $x_0=x_m$, we have replaced $u_0=u_m$ to determine the radius of the photon sphere. The behavior of the photon sphere radius is displayed in the top row (left panel) of Fig. \ref{fig8} versus the rotation parameter $a$ and taking different values for the electric charge $Q$. It can be observed, as both the rotation parameter $a$ and electric charge $Q$ decrease in magnitude, the radius of the photon sphere also decreases in magnitude. Therefore, the photon sphere size strongly depends upon the magnitude of the rotation parameter $a$ and electric charge $Q$. Moreover, the photon sphere increases to its maximum size for small electric charge $Q$ and counter-clockwise rotation, that is, when $a<0$. On the other hand, the impact parameter $u_m$ behavior is depicted on the right panel on the top row of Fig. \ref{fig8}. In this case, the situation is similar to that for the variable $x_m$, in particular, it decreases with increasing rotation parameter $a$ and electric charge $Q$.

Once the equations for the photon sphere radius and impact parameter are solved, the integral (\ref{angle}) can be expanded around $x_0=x_m$ to get more information about the deflection angle. To do this, it is necessary to introduce the following change of variable $z=1-x_0/x$ leading to
\begin{equation}\label{angleaprox}
\alpha_D(u)=-\bar{a}\text{Ln}\left(\frac{u}{u_m}-1\right)+\bar{b}+\mathcal{O}\left(u-u_m\right),
\end{equation}
where, $\{\bar{a},\bar{b}\}$ are the gravitational lensing coefficients given by
\begin{equation}
\bar{a}\equiv\frac{R(0,x_m)}{2\sqrt{c_{2m}}}, \quad \bar{b}\equiv-\pi+b_R+\bar{a}\text{Ln}\left(\frac{cx^2_m}{u^2_m}\right), 
\end{equation}
where, $b_R$ is the regular part of the integral (\ref{angle}) given by
\begin{equation}
b_R\equiv\int^1_0\left[R\left(z,x_m\right)f\left(z,x_m\right)-R\left(0,x_m\right)f_0\left(z,x_m\right)\right]dz,
\end{equation}
with
\begin{equation}
R\left(z,x_0\right)\equiv\frac{2x^2\sqrt{\bar{B}}\left(2\bar{A}_0\bar{A}u+\bar{A}_0\bar{D}\right)}{x_0\sqrt{\bar{C}\bar{A}_0}\sqrt{4\bar{A}\bar{C}+\bar{D}^2}}, \quad f\left(z,x_0\right)\equiv\frac{1}{\sqrt{\bar{A}_0+\bar{A}\frac{\bar{C}_0}{\bar{C}}+\frac{u}{\bar{C}}\left(\bar{A}\bar{D}_0-\bar{A}_0\bar{D}\right)}},
\end{equation}
with
\begin{equation}
f_0(z,x_0)\equiv\frac{1}{\sqrt{c_1z+c_2z^2}}.
\end{equation}
Here, the terms $c_1$ and $c_2$ are obtained by expanding the argument of the square root in $f(z,x_0)$ around $x=x_m$, and the coefficient $c$ is obtained by Taylor expansion of $u-u_m=c\left(x_0-x_m\right)^2$. As can be seen, the deflection angle (\ref{angleaprox}) blows up at $u=u_m$. This fact is observed from the middle row in Fig. \ref{fig8} (dots are representing the critical impact parameter where the deflection angle blows up). The left panel shows the trend of the deflection angle versus the impact parameter for $a<0$. As can be seen, the value $u_m$, where the divergence appears decreasing with decreasing impact parameter $u$ and electric charge $Q$. The same behavior is shown in the right panel for $a>0$, however, in this case, the different critical impact parameter values for the different values of the electric charge are closer than the previous case. In both cases, the deflection angle is super-passing $2\pi$, exhibiting a decreasing behavior with increasing impact parameter. In this concern, those values of the impact parameter drifting apart from the critical one are not relevant for SDL, because this technique is valid only in close vicinity to the critical impact parameter. Nevertheless, for each case near the critical values, the deflection angle becomes negative in nature. This fact is not acceptable in the BH case, since in such a case, this is related to a repulsive gravitational force. In the present case, as we are dealing with a wormhole, the presence of exotic matter at the throat and its neighborhood could in principle produce a repulsive gravitational interaction, making the path of the light or particles bend opposite to the wormhole throat, causing photons to be deflected away from the wormhole location. Although, this is not the case here, because our configuration contemplates the wormhole lying between a light source and observer, it is possible for a wormhole to obtain a complete negative deflection angle.

\subsection{Lens Equation and Observables}
To further understand the lensing phenomenon, it is essential to study the lens geometry, since in particular, we can identify the exact position and magnification of the relativistic images. As the light source and observer are placed far enough from the wormhole such that the gravitational forces around them are insufficient, the lens configuration we are interested concern position of the wormhole between a light source and observer \cite{Bozza:2002zjaaa,Bozza:2008evaaa}. Suppose that the source and observer are almost aligned, the lens equation reads
\begin{equation}\label{betabar}
\bar{\beta}=\theta-\frac{D_{LS}}{D_{OL}+D_{LS}}\Delta\alpha_n,
\end{equation}
where, instead of using a full deflection angle, we used an offset of deflection angle given by $\Delta\alpha_n=\alpha_D-2n\pi$, with $n\in\mathbb{N}$ and $0<\Delta\alpha_n\ll1$. Furthermore, $\beta$ and $\theta$ represent the angular positions of the source and image from the optical axis, respectively. On the other hand, the distances of the source and lens from the observer are given by $D_{LS}$ and $D_{OL}$, respectively. This information allows us to estimate some observables for the strong gravitational lensing \cite{Bozza:2002afaaa,Bozza:2002zjaaa,Bozza:2008evaaa}. Using the expressions (\ref{angleaprox}) and (\ref{betabar}), the angular separation between the lens and the $n^{\text{th}}$ image is given by \cite{Bozza:2002afaaa} 
\begin{equation}\label{thetan}
\theta_n=\theta^0_n+\Delta\theta_n,
\end{equation}
with,
\begin{equation}
\theta^0_n=\frac{u_m}{D_{OL}}\left(1+e_n\right), \qquad \Delta\theta_n=\frac{D_{OL}+D_{LS}}{D_{LS}}\frac{u_me_n}{\bar{a}D_{OL}}\left(\beta-\theta^0_n\right),
\end{equation}
such that
\begin{equation}
e_n=\exp\left(\frac{\bar{b}}{\bar{a}}-\frac{2n\pi}{\bar{a}}\right).
\end{equation}
Here $\theta^0_n$ is the angular position of the image when a photon encircles complete $2n\pi$ and $\Delta\theta_n$ is the extra term exceeding $2n\pi$ such that $\theta^0_n>>\Delta\theta_n$ \cite{Bozza:2002afaaa}. Now, considering a perfect alignment between the source and observer, that is $\bar{\beta}=0$, the lens deflects the light equally in all directions such that a ring-shaped image is produced, which is known as Einstein ring \cite{Bartelmann:1999yn}.

It should be noted that when the lens is exactly halfway between the observer and source, the angular radius of the Einstein rings can be obtained by solving Eq. (\ref{thetan}), for the source, lens and the observer being perfectly aligned \cite{Bozza:2002afaaa,Muller:2008zzg}, obtaining
\begin{equation}
\theta^{\text{E}}_n=\frac{u_m}{D_{OL}}\left(1+e_n\right).
\end{equation}
The value $n=1$ is the angular radius of the outermost Einstein ring. For the present wormhole model (\ref{32}), the outermost Einstein rings are illustrated in the bottom of Fig. \ref{8}. The left panel correspond to $a<0$ and the right one $a>0$. As can be observed, the ring increases its size as the electric charge $Q$ decreases in magnitude. On the other hand, in passing from $a<0$ to $a>0$, One can distinguish the rings much better, since they are further apart. Interestingly, in the gravitational lensing phenomenon, the light is deflected while maintaining the surface brightness, however, the appearance of the solid angle is continuously changing. This enhances the brightness of the images. The magnification for $n$-loop images is evaluated as the quotient of the solid angles subtended by the $n^{\text{th}}$ image and the source as follows \cite{Bozza:2002afaaa,Bozza:2002zjaaa} 
\begin{figure}[htbp]
\centering
\subfigure{\includegraphics[width=0.32\textwidth]{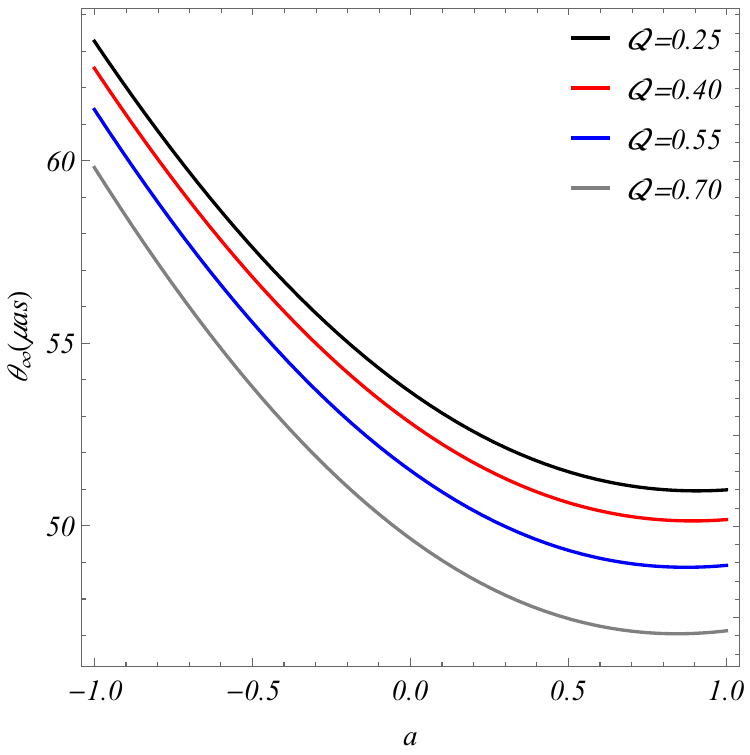}}~
\subfigure{\includegraphics[width=0.32\textwidth]{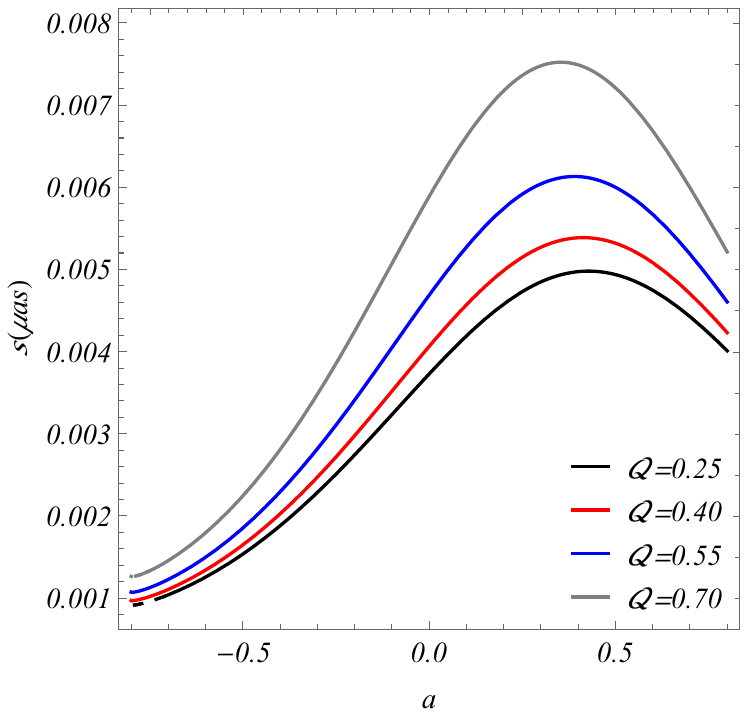}}~
\subfigure{\includegraphics[width=0.32\textwidth]{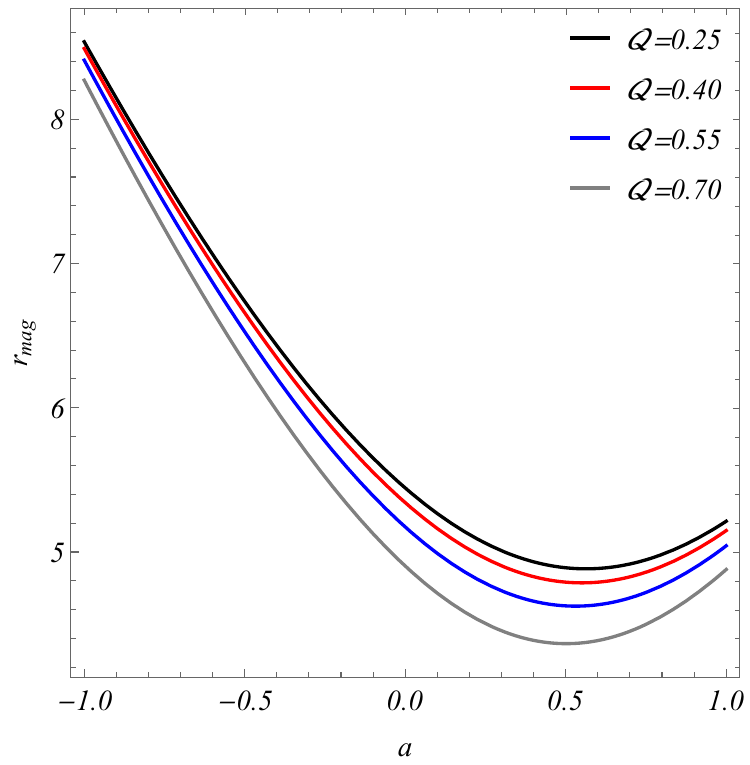}}~
\caption{The trend of the lensing observables $\theta_\infty$ (\textbf{left panel}), $s$ (\textbf{middle panel}) and $r_\text{mag}$ (\textbf{right panel}) versus the rotation parameter $a$. For all these plots we have considered $M=1$ and $b=0.5$. \label{fig9}}
\end{figure}
\begin{equation}\label{mag}
\mu_n=\frac{1}{\bar{\beta}}\left[\frac{u_m}{D_{OL}}\left(1+e_n\right)\left(\frac{D_{OL}+D_{LS}}{D_{LS}}\frac{u_me_n}{D_{OL}\bar{a}}\right)\right].
\end{equation}
Considering the case when the source and the observer are perfectly aligned ($\bar{\beta}=0$), the expression (\ref{mag}) diverges, suggesting that the perfect alignment maximizes the possibility of the detection of the images. Nevertheless, it is not an easy task to deal with it, then following \cite{Bozza:2002afaaa}, one reduces the complexity of the computations by considering the simplest situation in which only the outermost image, $\theta_{1}$, is resolved as a single image, while the subsequent images are all piled up together at $\theta_\infty$. Therefore, if 1-loop image can be distinguished from the rest of packed images, we can have three characteristic observables \cite{Bozza:2002afaaa,Bozza:2002zjaaa}, namely 
\begin{equation}
\theta_\infty=\frac{u_m}{D_{OL}}, \quad s=\theta_1-\theta_\infty\approx\theta_\infty\exp \left(\frac{\bar{b}}{\bar{a}}-\frac{2\pi}{\bar{a}}\right), \quad r_{\mathrm{mag}}=\frac{\mu_1}{\sum_{n=2}^{\infty}\mu_n}\approx\frac{5\pi}{\bar{a}\log(10)}.
\end{equation}
In the above expression, $\theta_{1}$ is the angular position of the outermost single image, $\theta_\infty$ is the angular position of the rest of packed images, $s$ is the angular separation between the $\theta_{1}$ and $\theta_\infty$, $r_\text{mag}$ is the ratio of the flux of the first image and the all other images. In Fig. \ref{fig9}, we have plotted the trend of the lensing observables $\theta_\infty$ (left panel), $s$ (middle panel) and $r_\text{mag}$ (right panel) versus the rotation parameter $a$ and different values of the electric charge $Q$. It is clear from the panel exhibiting in Fig. \ref{fig9} that all the lensing observables are attaining their maximum values for $a<0$ and small electric charge $Q$.

\section{Concluding Remarks}\label{sec6}
In this work, using the algorithm given in \cite{2016EPJC...76....7Aaaa}, a new rotating wormhole solution has been obtained starting from the static wormhole spacetime provided in \cite{2023AnPhy.45769411Faaa}. This solution is sourced by a non-conventional electrodynamics, the so-called Bopp-Podolsky electrodynamics \cite{Boppaaa,PhysRev.62.68aaa}. The rotating wormhole model given in the Kerr-like form (\ref{31}), is consistent with the static limit, that is, when $a\rightarrow 0$ the original static solution is recovered. Moreover, taking the limit $b\rightarrow 0$ in the rotating version (\ref{31}), one recasts the well-known Kerr-Newman BH and taking both $Q\rightarrow0$ and $b\rightarrow 0$ the Kerr BH is recovered. Therefore, non-trivial charged rotating wormhole in the Bopp-Podolsky electrodynamics scenario are possible only if $Q\neq 0$ and $b\neq 0$. However, in this case, the main geometric features that any wormhole topology must satisfy, are not guarantee for all $(Q,b)$ pair (for a given mass $M$ and rotation parameter $a$). In this concern, we have performed a thorough analysis about the possible combinations of $(Q,b)$ to have a well-defined wormhole geometry. This analysis reveals that for increasing $Q$ and $b$, the wormhole throat increases it size.

To further validate the feasibility of this theoretical wormhole model, we performed a detailed analysis about some optical properties, such as the shadow and the strong gravitational lensing. The former was performed using the well-known Hamilton-Jacobi approach, where the null geodesics are determined by the technique of separation of variables in $r$ and $\theta$ coordinates. Thus, allowing us to study the effective potential and the behavior of unstable photon orbits. Keeping a simple scenario in the equatorial region, we computed the effective potential with respect to radial distance $r$. It was found that the spin and charge had identical impact on the variation and behavior of the effective potential and the unstable orbits. That is, with increase in $Q$ and $a$, the effective potential increased, while the unstable orbits shifted towards the throat. Contrary to this behavior, with increase in parameter $b$, the effective potential decreased and the unstable orbits shifted away from the throat. Further, we found the shadow images for the prescribed parametric values. These images are found to be identical to those for the BHs which is due to the fact that the parametric values of $a$ are kept sufficiently smaller than its critical values above which no shadow is formed. The shadows became smaller with respect to $Q$ and shifted towards the right with respect to $a$. However, a significant variation in shadows was observed with respect to $b$ only for larger values of either $a$ or $Q$. For critical and near critical spin values, it was found that the inner and outer unstable orbits did not merge smoothly and the formed shadow curves were cuspy as displayed in Figs. \ref{Csp1} and \ref{Csp2}.

Regarding the strong gravitational lensing study, here we have considered the case where the observer and source are on opposite side of the wormhole throat. Using Bozza's procedure \cite{Bozza:2002afaaa,Bozza:2002zjaaa,Bozza:2008evaaa} the deflection angle has been determined and depicted in Fig. \ref{fig8} versus the impact parameter $u_{m}$, where the incidence of the electric charge and rotating parameter on its behavior is relevant. In this concern, a small amount of electric charge induces divergence of the deflection angle for smaller values of the critical impact parameter. Besides, the size of Einstein rings is also reduced for larger values of the electric charge. Additionally, for the lensing observables, such as angular position of the formed images, their angular separation and ratio of the flux of the first image with respect to the all other images have been explored. Interestingly, the angular separation between the outermost and asymptotic relativistic images increased as the electric charge increases in magnitude (see middle panel in Fig. \ref{fig9}) and the relative magnification of the outermost relativistic image with other relativistic images decreased with increasing electric charge (see right panel in Fig. \ref{fig9}). It is clear that in considering the configuration taken here to study strong gravitational lensing phenomenon, in principle will be quite difficult to recognize if the observed object is corresponding to a rotating BH or to a rotating wormhole.

Finally, we can remark that wormholes are considered to be one of the mysteries of the universe that are yet to be discovered observationally to prove their existence. The concept of light deflection and trapping of photons in orbits around a wormhole is quite similar to that in the case of a BH. However, depending upon the structure and geometry of wormholes, it is hard to visualize and understand the shadow and lensing patterns formed by the deflection of light. In this concern, it is clear that optical properties (at least from the theoretical point of view) behave similarly as they do for BHs, then it is difficult to visualize from this perspective the main difference between these objects. Therefore, this implies that further studies are necessary to fully constrain and realize the relevant observational parameter to distinguish these structures from each other.

\section*{Acknowledgments}
The authors are thankful to Mustapha Azreg-Ainou for his valuable suggestions on the construction of the rotating wormhole metric and the reduction of the metric to the Morris-Thorne form. The work of M. Zubair has been partially supported by the National Natural Science Foundation of China under project No. 11988101. He is grateful to compact object and
diffused medium Research Group at NAOC led by Prof. JinLin Han for excellent hospitality and friendly environment.

\end{document}